\documentclass[preprint]{aastex}

\usepackage{graphicx}
\usepackage{amsmath,amsfonts,amssymb}
\usepackage[T1]{fontenc}
\usepackage{txfonts}
\usepackage{natbib}

\shorttitle{A stellar pulse imaged by optical interferometry}
\shortauthors{S. Lacour et al.}

\begin{document}
   \title{The Pulsation of $\chi$ Cygni Imaged by Optical
     Interferometry;\\ a Novel Technique to Derive Distance and Mass
     of Mira Stars}

\author{S. Lacour\altaffilmark{1},
 E. Thi\'ebaut\altaffilmark{2},
 G. Perrin\altaffilmark{1},
 S. Meimon\altaffilmark{3},
 X. Haubois\altaffilmark{1},
 E. Pedretti\altaffilmark{4},
 S. Ridgway\altaffilmark{5}, \\
 J.D. Monnier\altaffilmark{6}, 
 J.P. Berger\altaffilmark{7},
 P.A. Schuller\altaffilmark{8}, 
 H. Woodruff\altaffilmark{9}, 
 A. Poncelet\altaffilmark{1}, 
 H. Le Coroller\altaffilmark{10}, \\
 R. Millan-Gabet\altaffilmark{11}, 
 M. Lacasse\altaffilmark{12}, 
and W. Traub\altaffilmark{13} 
}


\altaffiltext{1}{ Observatoire de Paris, LESIA, CNRS/UMR 8109, 92190 Meudon, France}
\altaffiltext{2}{ Centre de Recherche Astrophysique de Lyon, CNRS/UMR 5574, 69561 Saint Genis Laval, France}
\altaffiltext{3}{ Office National d'\'Etudes et de Recherches A\'eronautiques, DOTA, 92322 Chatillon, France}
\altaffiltext{4}{ School of Physics and Astronomy, University of St. Andrews, North Haugh, St Andrews KY16 9SS, United Kingdom}
\altaffiltext{5}{ National Optical Astronomy Observatory, P.O. Box 26732, Tucson, AZ 85726-6732, USA}
\altaffiltext{6}{ University of Michigan, Astronomy dept., 914 Dennison bldg., 500 Church street, Ann Arbor, MI, 40109, USA}
\altaffiltext{7}{ LAOG-UMR 5571, CNRS and Universit\'e Joseph Fourier, BP 53, 38041 Grenoble, France}
\altaffiltext{8}{ Institut d'Astrophysique Spatiale, CNRS/UMR 8617, Universit\'e Paris-Sud, 91405 Orsay, France}
\altaffiltext{9}{ Sydney Institute for Astronomy (SIfA), School of Physics, University of Sydney, NSW 2006, Australia}
\altaffiltext{10}{ Observatoire de Haute-Provence, OHP/CNRS, F-04870 St. Michel l'Observatoire, France     }
\altaffiltext{11}{Michelson Science Center, California Institute of Technology, MS 100-22, Pasadena, CA 91125, USA}
\altaffiltext{12}{ Harvard-Smithsonian Center for Astrophysics, 60 Garden Street, Cambridge, MA, 02138, USA}
\altaffiltext{13}{ Jet Propulsion Laboratory, California Institute of Technology, M/S 301-451, 4800 Oak Grove Drive, Pasadena, CA, 91109, USA
}


 
\begin{abstract}
We present infrared interferometric imaging of the S-type Mira star
$\chi$ Cygni. The object was observed at four different epochs in
2005-2006 with the IOTA optical interferometer (H\ band). Images show
up to $40\%$ variation in the stellar diameter, as well as significant
changes in the limb darkening and stellar inhomogeneities. Model
fitting gave precise time-dependent values of the stellar diameter,
and reveals presence and displacement of a warm molecular layer.  The
star radius, corrected for limb darkening, has a mean value of
$12.1\,$mas and shows a $5.1\,$mas amplitude pulsation.  Minimum
diameter was observed at phase $0.94\pm0.01$. Maximum temperature was
observed several days later at phase $1.02\pm0.02$.  We also show that
combining the angular acceleration of the molecular layer with CO
($\Delta v = 3$) radial velocity measurements yields a $5.9\pm1.5\,$
mas parallax. The constant acceleration of the CO molecules -- during
80\% of the pulsation cycle -- lead us to argument for a free-falling
layer. The acceleration is compatible with a gravitational field
produced by a $2.1^{+1.5}_{-0.7}$ solar mass star. This last value is
in agreement with fundamental mode pulsator models.  We foresee
increased development of techniques consisting in combining radial
velocity with interferometric angular measurements, ultimately
allowing total mapping of the speed, density, and position of the
diverse species in pulsation driven atmospheres.
\end{abstract}


%

\section{Introduction}

Mira variables are low to intermediate mass AGB stars that pulsate
with a period of about 1 year. They have a cool ($T_{\rm eff}\leq
3000\,K$) and extended ($R>100\,R_\sun$) photosphere. As such, they are
bright ($M_k \leq -7$) infrared beacons, individually observable far
into galaxies of the Local Group \citep{1999IAUS..191..551Z}. They
have the potential to probe places where the distance \citep[eg,
  NGC\,5128 in][]{2004A&A...413..903R} or reddening \citep[eg, the
  Galactic Center in][]{2009MNRAS.tmp.1192M} does not allow
observation of the fainter/bluer -- and rarer -- Cepheids.

However, the challenge to overcome is that Mira stars are both
intrinsically complicated and ill-understood.  Two important relations
are of special interest: the period/luminosity (P/L) and the
period/mass/radius (P/M/R).  The first relation has been derived from
population studies (sequence ``C'' in the LMC from
\citet{2000PASA...17...18W} and also in the globular cluster 47 Tuc
from \citet{2005A&A...441.1117L}).  The present best parameterization
of the P-L relation within our galaxy is \citep{2008MNRAS.386..313W}:
\begin{equation}
M_k=-(3.51\pm0.20)(\log(P)-2.38)-(7.25\pm0.07)\,,
\label{eq:PL}
\end{equation}  
where $P$ is the period in days.  The zero
point of this relation is the most uncertain parameter, with its
dependence on the metallicity hardly known. The main difficulty is
that parallax values are inaccurate and error-prone due to the large
size and inhomogeneous surface brightness of the objects. 

The second relation, the P/M/R relation, has more relevance to the
fundamental physics of the star. It is extremely dependent on the
pulsation mode, but also, less crucially, on the surface density and
metallicity. The P/M/R relation has been formally derived from
numerical modeling of the fundamental pulsation mode of theses stars
\citep{Wood..89}:
\begin{equation}
\log(P)=-2.07+1.94 \log(R/R_\sun)-0.9 \log(M/M_\sun)\,,
\label{eq:PMR}
\end{equation}
Twenty years later, this model-derived relationship has still rarely
been confronted with observation. This paper is a first step forward
to establish the P/L and P/M/R relations on a new firm
observational footing.

Of the three crucial parameters (distance, mass, and radius), the
angular diameter is far from being the easiest value to obtain.
Because the surface gravity is several orders of magnitude lower
than the sun, the pulsation of Mira variable leads to an extended
 atmosphere. In the cool upper layers, significant amount of
the products of the helium fusion react to form di- and polyatomic
molecules including TiO, SiO, CO and H$_2$O.  The forest of molecular
lines and scattering from the dust lead to exotic intensity
distributions not at all like a simple stellar disk. In the past, this
substantially affected many stellar angular diameters measurements
\citep{1996AJ....112.2147V,1999A&A...345..221P}, paving the way to
contentious discussion on the mode of pulsation
\citep{1998A&A...333..647B,1999ApJ...514L..35Y}.


However, nowadays, interferometers are able to provide maps of the
brightness distribution as a function of wavelength
\citep{2009A&A...496L...1L,2009ApJ...700..114P}. Images of the Mira
star T\ Lep revealed a shell-like atmosphere, with a bright chromatic
zone distinctly detached from the photosphere. This could be the first
image of what \citet{2004A&A...424.1011O} called the MOLsphere, a zone
of increased density in which formation of warm molecular species
would be favored. Accounting for this layer is the key to obtain a
correct value for the diameter
\citep{1999A&A...345..221P,2002ApJ...579..446M,2004A&A...426..279P}. We
will also show that we can apply to this layer a modified
Baade-Wesselink method to derive the distance and mass of the star.

The test star of this paper is $\chi$ Cyg, a S-type Mira star. It
has a pulsation period of 408 days, a photometric magnitude ranging
from 5.3 to 13.3, and intense emission lines at postmaximum
\citep{1947ApJ...106..274M}. This suggests a large pulsation
amplitude.  Images were obtained with the IOTA interferometer at four
different stellar phases, chronologically $\phi = 0.93$, 0.26, 0.69
and 0.79. In the next section, we describe the observations and give a
short overview of the dataset. In section~\ref{sc:imaging}, we use an
image reconstruction algorithm to map the brightness distribution of
the star. Precise geometrical parameters of the star, including
existence of the molecular layer, are determined by model fitting in
section~\ref{sc:param_image}. From these values, temperatures and
opacities are deduced in section~\ref{sec:Physics}. Finally, in
section~\ref{sc:der_ma}, we combine angular acceleration with radial
velocities measurements to derive estimations of the distance and mass
of the star.


\section{Observation and data reduction}

The interferometric data presented herein were obtained using the IOTA
(Infrared-Optical Telescope Array) interferometer
\citep{2003SPIE.4838...45T}, a long baseline interferometer which
operated at near-infrared wavelengths. It consisted of three
0.45-meter telescopes movable among 17 stations along two orthogonal
linear arms. IOTA synthesized a total aperture size of $35 \times
15\,$m, corresponding to an angular resolution of $\approx 10 \times
23$ milliarcseconds at 1.65 $\mu$m.  Squared visibility and closure phase
measurements were obtained using the integrated optics combiner IONIC
\citep{2003SPIE.4838.1099B}. IOTA ceased operation in July 2006.

\begin{figure}
   \centering
\includegraphics[width=7.5cm]{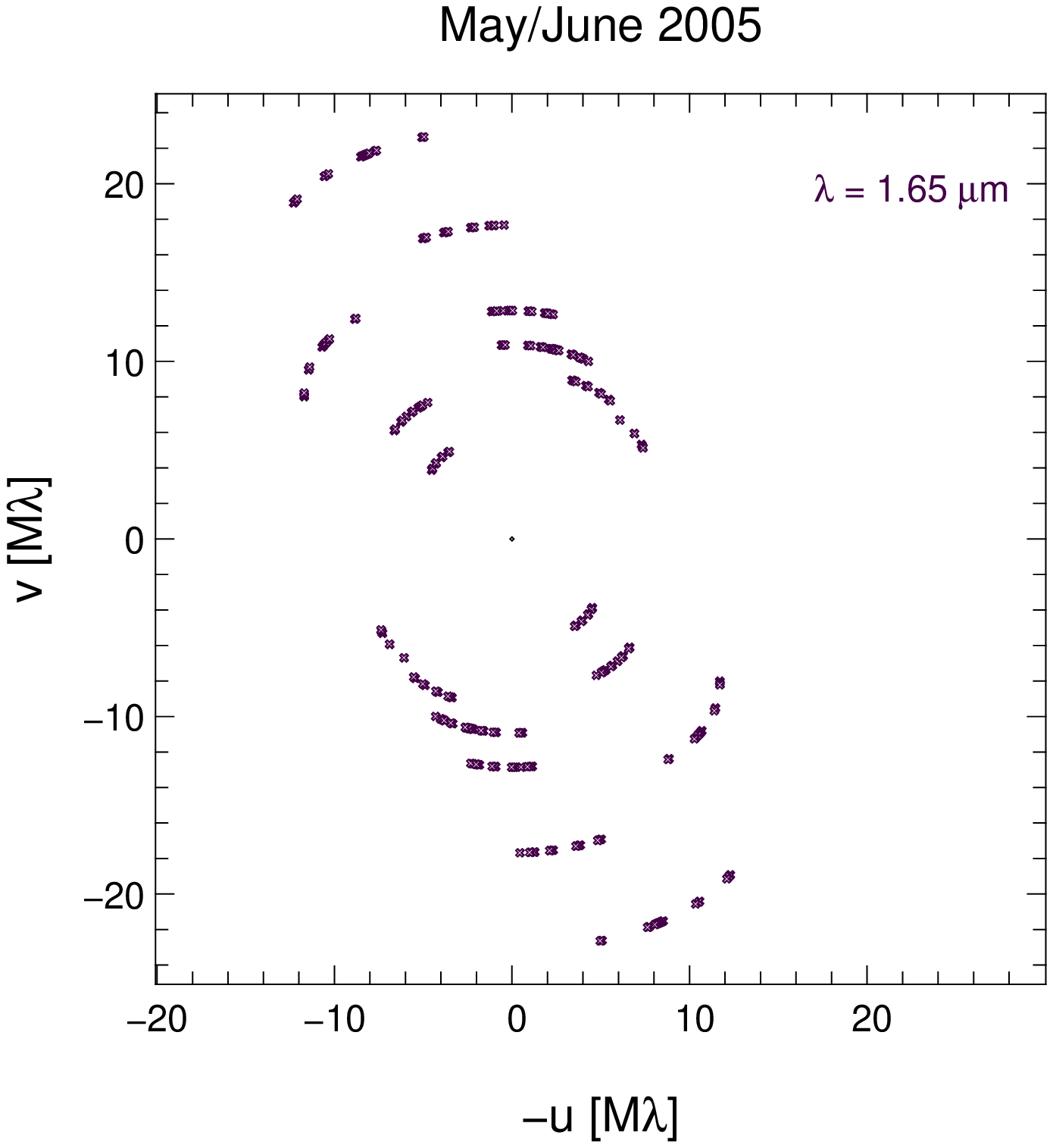} 
\includegraphics[width=7.5cm]{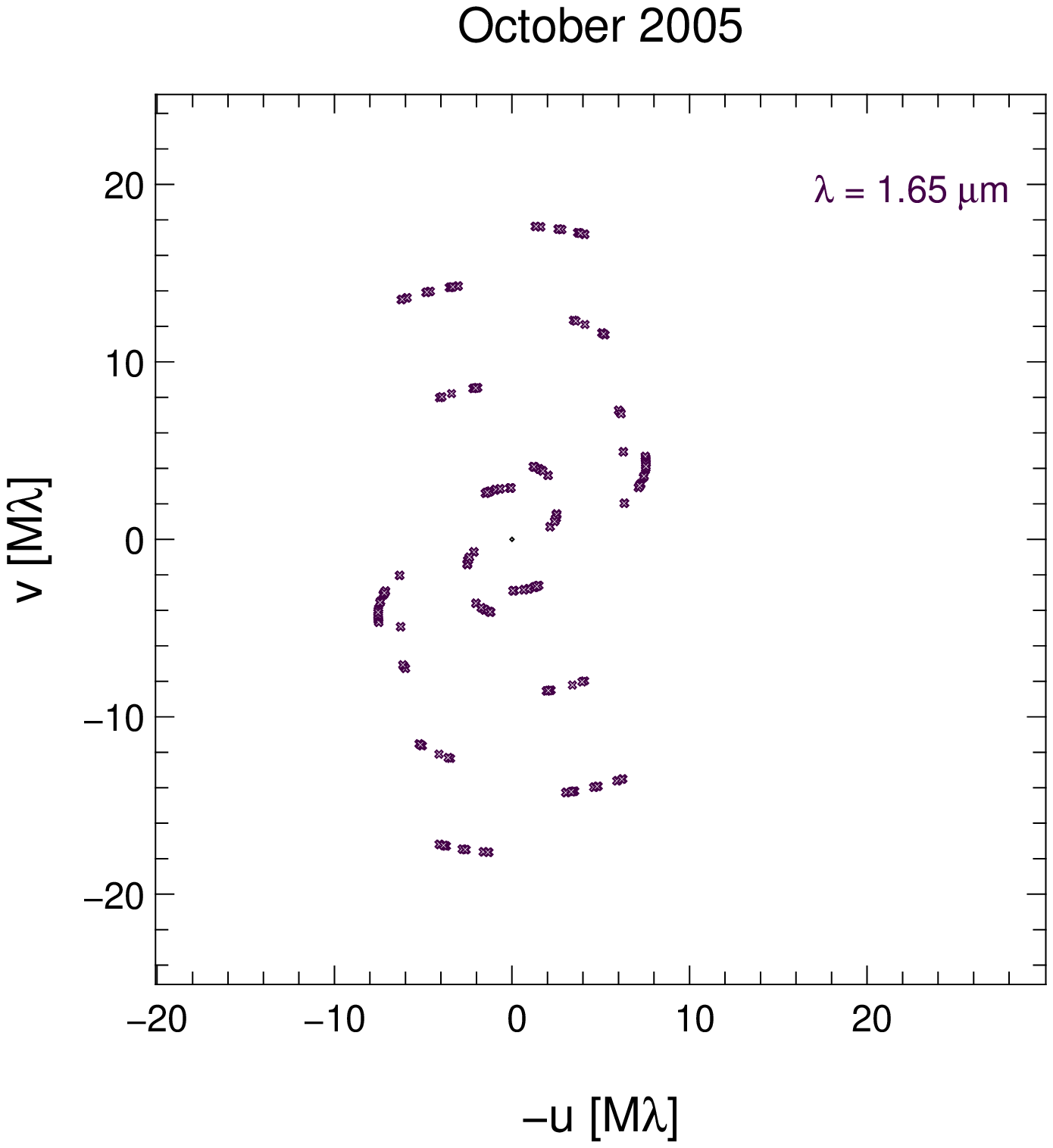}\\ \vspace{.1cm}
\includegraphics[width=7.5cm]{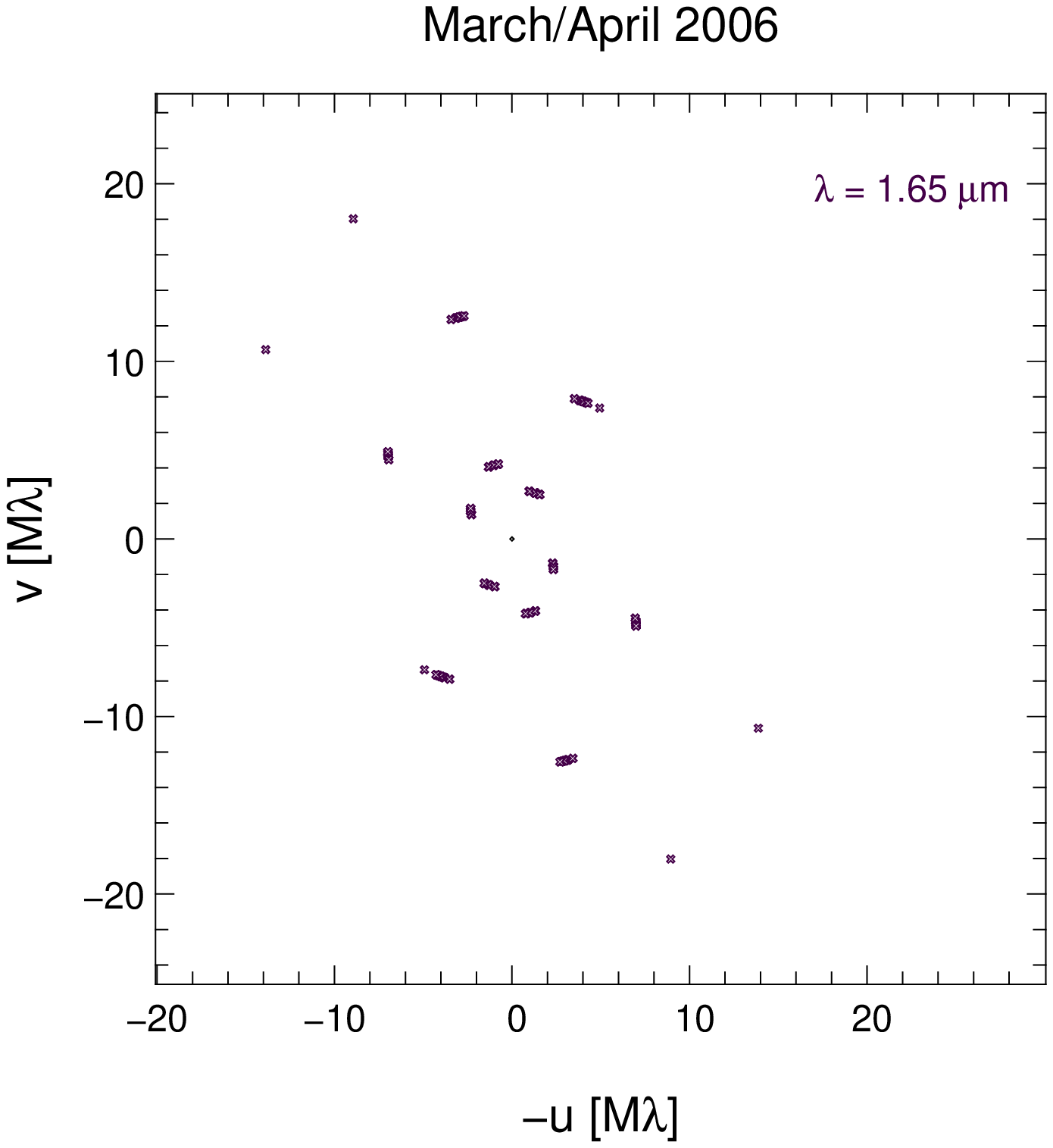} 
\includegraphics[width=7.5cm]{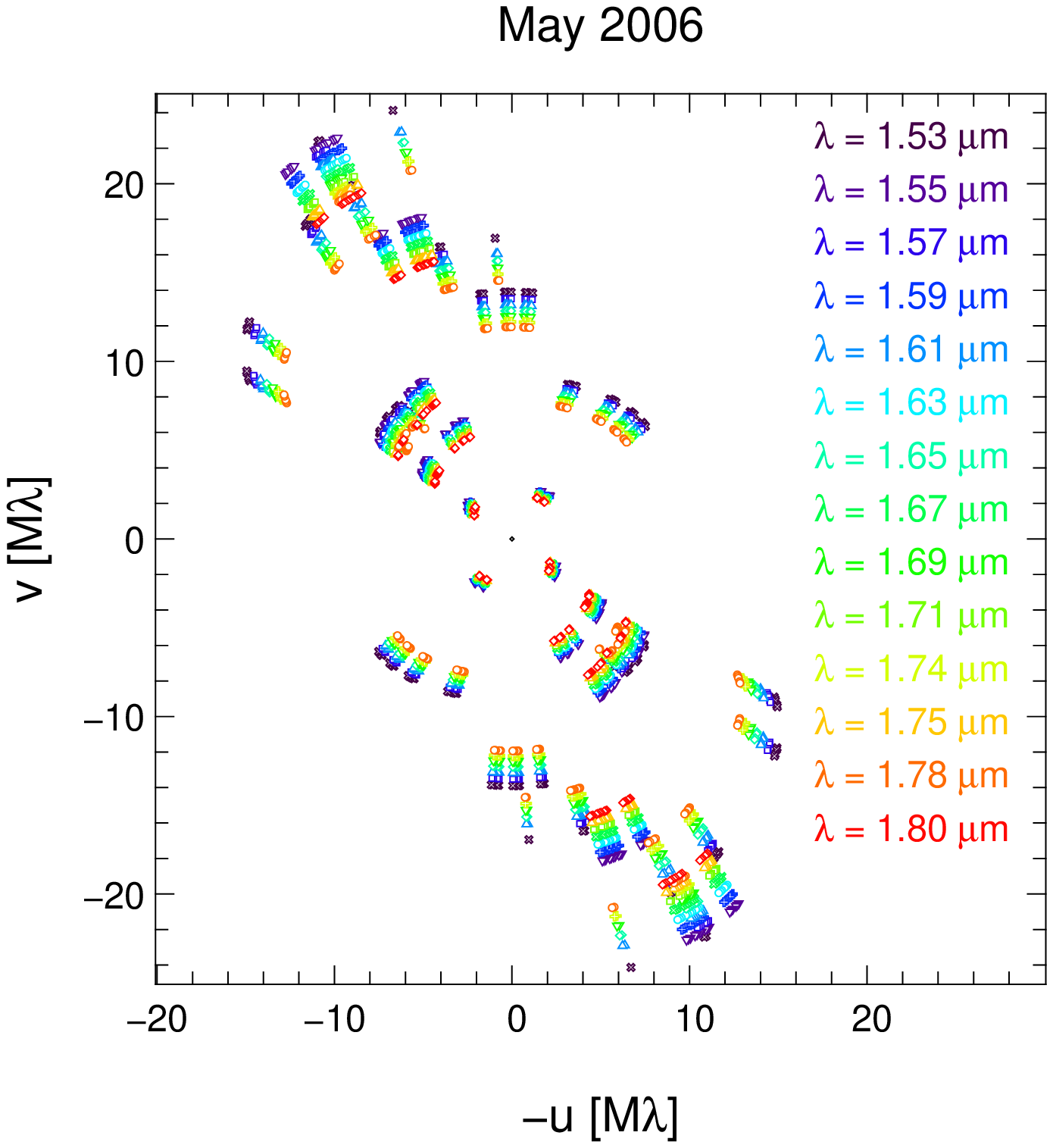}
      \caption{$u$-$v$ coverage obtained on $\chi\,$Cyg at the four
        epochs of observation. Coordinates are in factors of the
        wavelength, ie, 10\,M$\lambda$ correspond to a baseline of 16
        meters at $\lambda=1.6\,\mu$m. Baseline lengths range from 5
        meters to 38 meters.  }
         \label{fig:UV_planes}
   \end{figure}

The declination of $\chi$ Cyg made possible observations at low
airmass with the IOTA interferometer during 6 months per year. With a
period of 408 days, it allowed observations around both the minimum
and the maximum brightness.  It was observed at four different stellar
phases: May 2005 ($\phi = 0.93$), October 2005 ($\phi = 0.26$), April
2006 ($\phi = 0.69$) and May 2006 ($\phi = 0.79$).  Observation
information can be found in Table~\ref{tb:Config}, including dates and
phase of observation, interferometer configurations and projected
baseline lengths.  Fig.~\ref{fig:UV_planes} shows the $u$-$v$ coverage
achieved during these observation runs. The geometry of the IOTA
interferometer and the position of the star on the sky constrained the
extent of frequency coverage.

The first three periods of observations were undertaken with a narrow
H band filter ($1.59\,\mu$m$\,\leq\,\lambda\leq\,1.69\,\mu$m). A
dispersive setup was implemented shortly before our last observation
run, resulting in a dataset featuring 7 spectral channels covering the
range $1.5\,\mu$m$\,\leq\,\lambda\leq\,1.8\,\mu$m
\citep{2008SPIE.7013E..88P}.  The science target observations are
interleaved with identical observations of unresolved or partially
resolved stars, used to calibrate the interferometer's instrumental
response. The interleaved calibrator sources (listed in
Table~\ref{tb:calib}) were chosen in two different catalogs:
\citet{2002A&A...393..183B} and \citet{2006A&A...447..783M}.

Reduction of the IONIC3 data was carried out using custom
software \citep{2006ApJ...647..444M}, with exactly the same settings
as the ones previously detailed in \citet{2008A&A...485..561L}. The
output of this reduction software are squared visibilities ($V^2$) and
closure phases (CP). Data are available upon request in the OIFITS
format \citep{2005PASP..117.1255P}.  They are presented in the 8
panels of Fig.~\ref{fig:Res_Vis1} and~\ref{fig:Res_Vis2} (superimposed
on the data are curves of the best fit of the model presented in
Sect.~\ref{sc:param_image}).  Eye analysis of the $V^2$ (upper panels)
show a significant diameter variation between the different epochs.
Also, at the longest baseline, the CP are clearly different from 0 or
$\pi$. This reveals the presence of an asymmetric brightness
repartition, even though the complexity of CP estimators makes it hard
to straightforwardly derive the level of asymmetry.

\section{Regularized imaging} \label{sc:imaging}

A first step in our data analysis was to convey the squared visibilities and
closure phase into spatial information. 
The imaging was performed by the \textsc{Mira} reconstruction
software\footnote{
  \url{http://www-obs.univ-lyon1.fr/labo/perso/eric.thiebaut/mira.html}}.
It stands for ``Multi-aperture Image Reconstruction Algorithm''
\citep{2008SPIE.7013E..43T,Thiebaut_etal-2003-JENAM}. For more
details, we refer the readers to previous image reconstruction work
using this software \citep{2008A&A...485..561L,2009A&A...496L...1L} as
well as more theoretical descriptions
\citep{2008ISTSP...2..767L,2009ISTSP...Thi}.

Because the frequency coverage is far from being complete, the
theoretical bijection between the frequency (Fig.~\ref{fig:UV_planes})
and spatial domains is hard to achieve. It explains why the imaging
algorithm requires a strong regularization term: the image is sought
by minimizing a so-called \emph{cost function} which is the sum of a
regularization term plus data related terms. The data terms enforce
the agreement of the model image with the different kind of measured
data (squared visibilities and phase closures).  The interpolation of
missing frequency coverage is performed by the regularization and by
strict constraints such as the positivity (which plays the role of a
floating support constraint) and normalization.

For $\chi$ Cygni, the regularization term was a $\chi^2$ minimization
between the reconstructed image and a simple model of limb-darkened
disk. Explicitly, the regularization term was:
$\alpha\,\sum_k(l_k-m_k)^2$ where $\alpha$ is the relative weight of
the regularization, $l_k$ is the intensity of $k$-th pixel in the
sought image and $m_k$ is the flux of the $k$-th pixel given by the
limb darkening model. The $m_k$ is a model of the brightness
distribution of a stellar disk whose parameters (diameter and power of
the limb-darkening) are beforehand adjusted on the squared
visibilities. An advantage of this image reconstruction approach is
that our prior favor radial symmetry. Hence this gives more strength
to the relevancy of asymmetric features such as brighter regions in
the restored images.

\begin{figure}
   \centering
   \includegraphics[width=15cm]{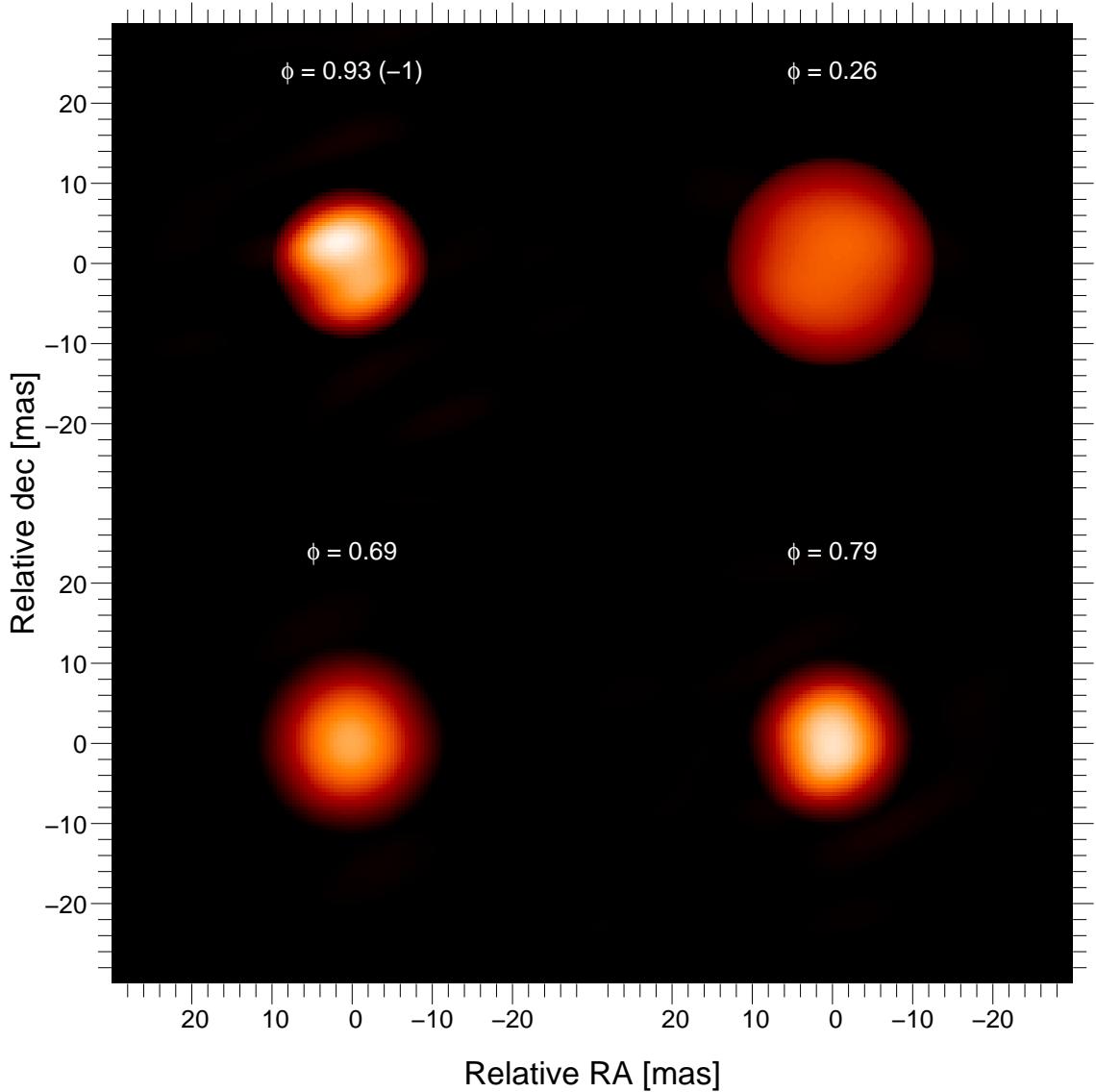}
      \caption{Regularized imaging of $\chi$ Cyg. The four epochs are
        labeled by their stellar phase, from upper left hand to lower
        right hand. The variation in diameter is eye striking with
        diameter changes of up to 40\% between phase 0.93 and
        0.26. Changes in limb darkening are also present.  Hot cells
        on the photosphere are apparent, interestingly with a higher
        contrast of the cells when the star is at smaller
        diameter. The relative brightnesses are normalized according
        to H band magnitude from Table~\ref{tb:flux}.  The angular
        resolution of the IOTA interferometer is $10\times23\,$mas
        ($1.65\mu$m).  At $\chi$ Cygni's distance of 170\,pc, 5.9\,mas
        correspond to 1\,AU (Sect.~\ref{sc:para}).  }
         \label{fig:Reg_Image}
   \end{figure}

The results are presented in Fig.~\ref{fig:Reg_Image}. The four epochs
are labeled by their stellar phase from upper left hand to lower right
hand. The color scales of each image reconstruction is the same. The
total brightness of each epoch is normalized according to the
bolometric flux estimated in Sect.~\ref{sc:eff_temp}.
 
\begin{figure}
   \centering
   \includegraphics[width=13cm]{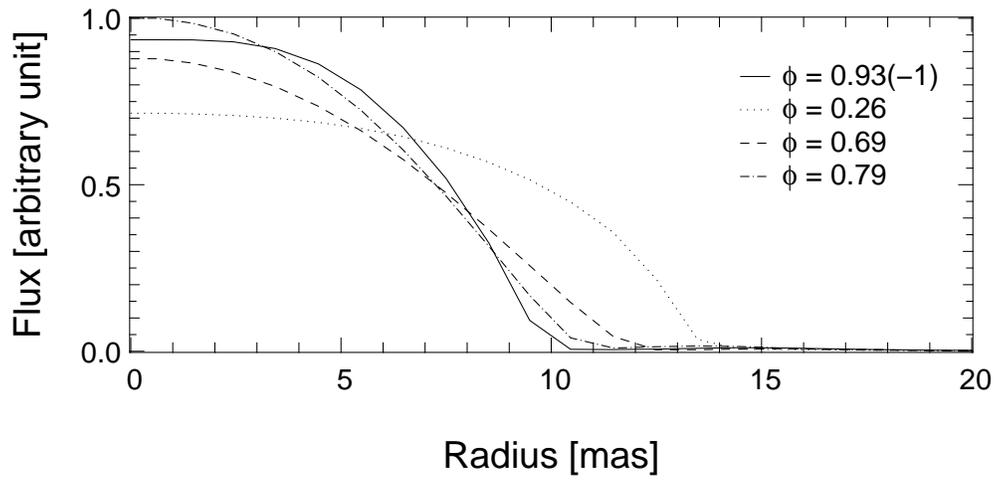}
      \caption{Center-to-limb variation (CLV) at the four epochs of
        Fig~\ref{fig:Reg_Image}. The flux is radially averaged. It
        shows a clear variation in the brightness distribution as a
        function of time.}  \label{fig:Reg_Image_r}
   \end{figure}

The variation in diameter is eye striking. Radius (defined by a 10\%
of the maximum surface flux) range from 9.4 ($\phi=0.93$) to 13.2\,mas
($\phi=0.26$). This means a 40\% expansion of the photosphere between
the two epochs. Curves in Fig.~\ref{fig:Reg_Image_r} represent the
radially averaged flux as a function of the radius. It is clear that,
on top of diameter variations, changes in limb darkening are present,
making it an important factor which has to be accounted for to
determine precise photospheric radii.

\section{Model fitting} \label{sc:param_image}

\subsection{Choosing the right geometrical description} \label{sc:param}

\begin{figure}
   \centering
   \includegraphics[width=15cm]{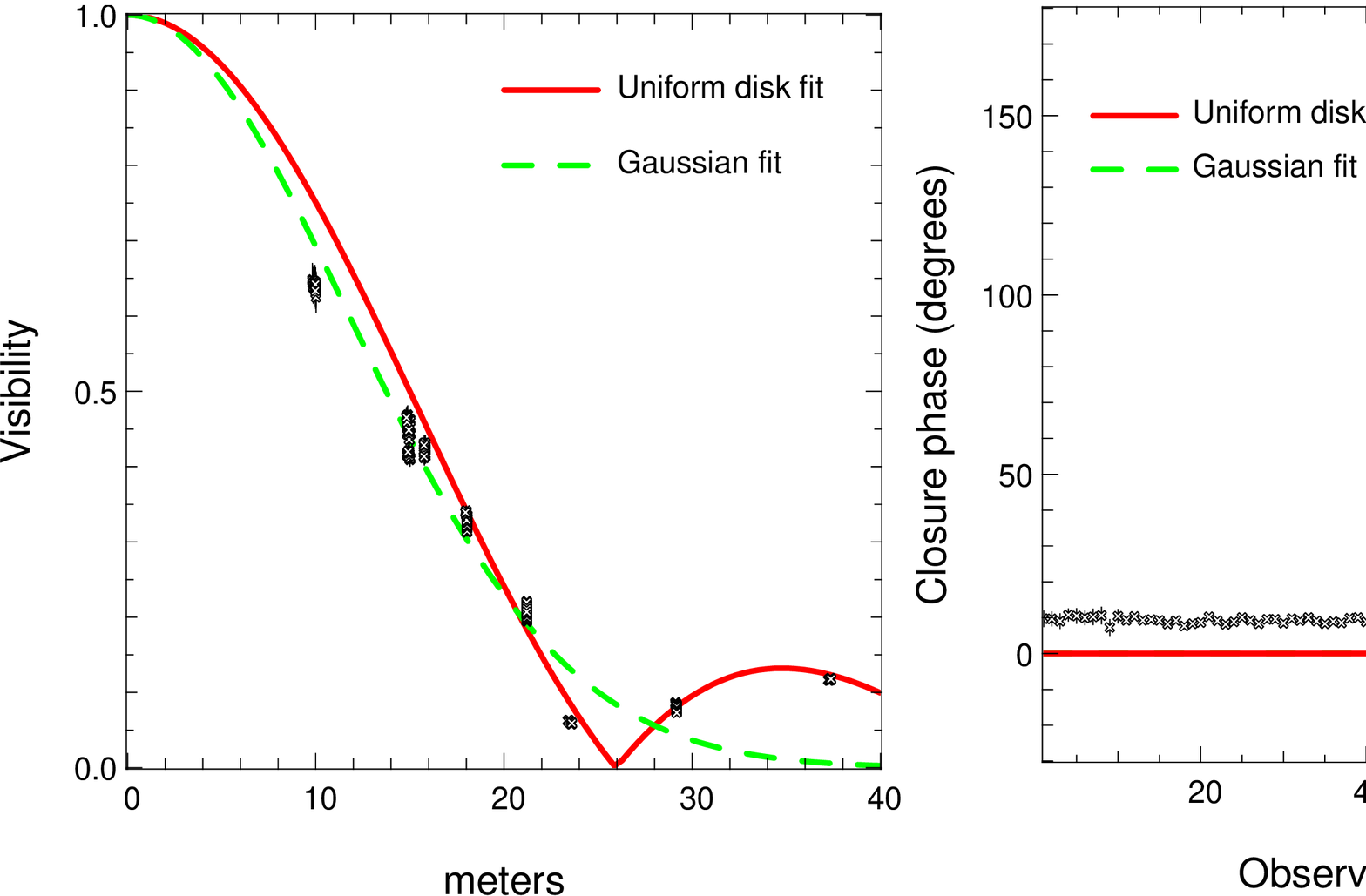} \\
   \includegraphics[width=15cm]{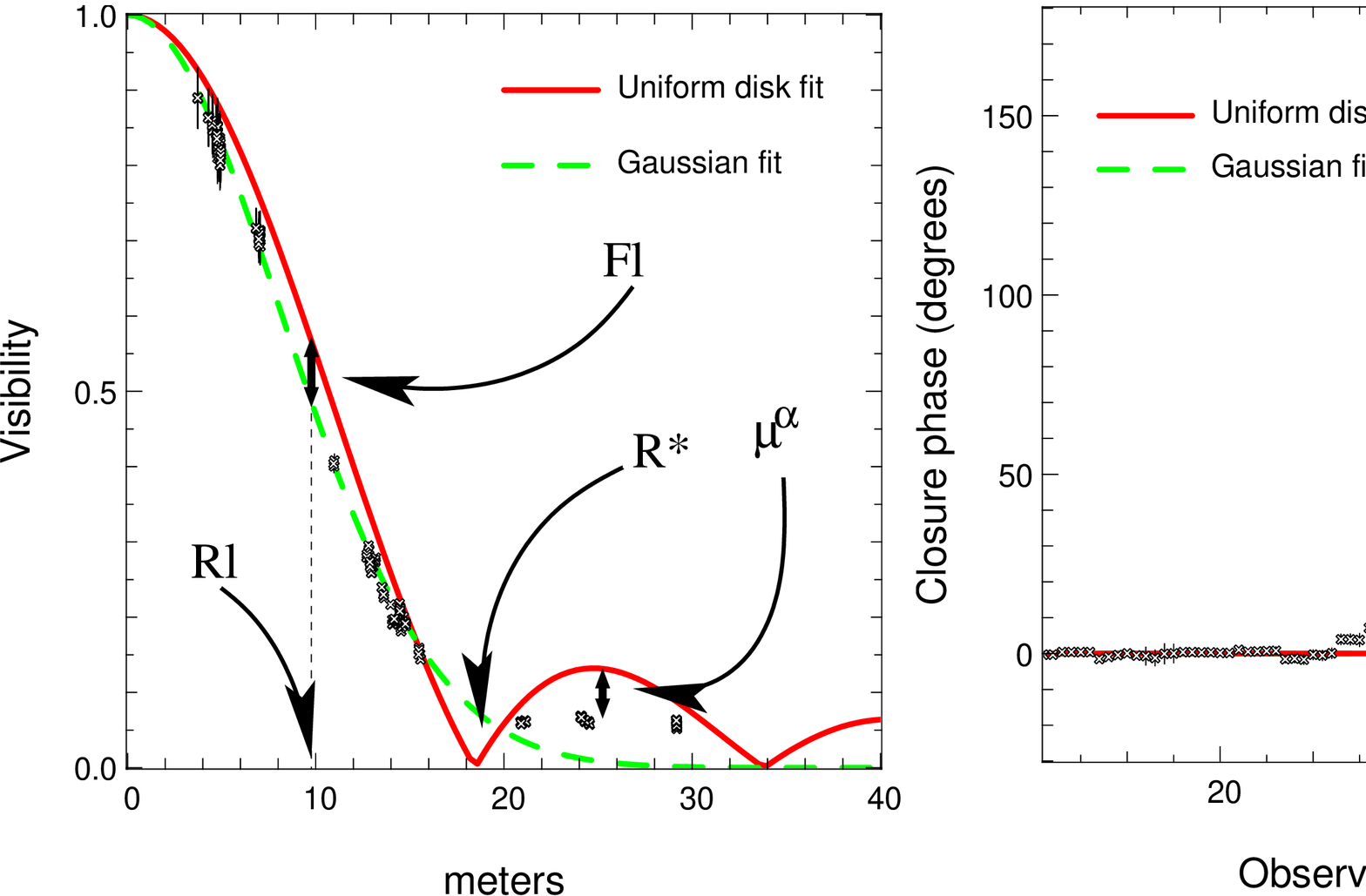}
      \caption{ Data of May 2005 (upper panels) and October 2005
        (lower panels). On top of the measurements are plotted the
        best fit of a uniform disk and a Gaussian disk. While the UD
        has trouble to fit the low frequency measurements, the GD is
        unable to account for the CP $\pi$-shift.  In the two lower
        panels are also summarized the ``zone of influence'' of the
        parameters used in the model presented in
        Fig.~\ref{fig:model}: the molecular layer (of flux $F_l$ and
        radius $R_l$) will impact the data at low frequency, the
        diameter of the star ($R_\star$) will determine the position
        of the first null, and the limb darkening ($\mu^\alpha$) will
        impact the higher frequency by modifying the height of the
        second lob. The asymmetry is clearly revealed by the CP.}
         \label{fig:UDUG}
   \end{figure}

Interferometric observation of Mira stars are traditionally
interpreted by fitting simple brightness profiles, such as a uniform
disk (UD) or a Gaussian disk (GD) \citep{2003SPIE.4838..163S}. Both
fits showed inadequacy with our data. This is shown at the two first
epochs of observation on Fig.~\ref {fig:UDUG}. At low frequency, the
data are best fitted with a Gaussian disk. At higher frequency,
however, closure phases $\pi$-shifted imply the presence of a second
lobe, ruling out a Gaussian type profile. With reduced $\chi^2$ well
over several hundreds, these fits confirm the need for a more complex
geometrical description.

\begin{figure}
   \centering
   \includegraphics[width=7.5cm]{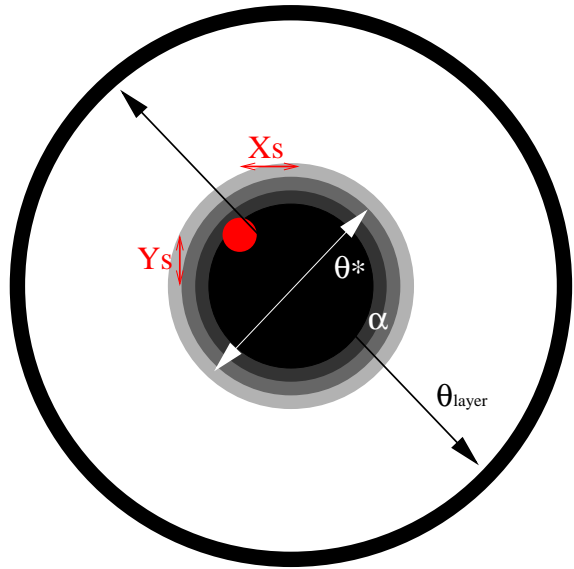}
      \caption{The model used in Sect.~\ref{sc:param_image}. It
        consists of a limb-darkened disk, a spot and a molecular layer
        represented by a ring. The free parameters are:
        $\theta_\star$, $\theta_{\rm layer}$, $\alpha$, $X_s$, $Y_s$,
        plus the brightness of the layer and the spot relative to the
        total flux.}
         \label{fig:model}
   \end{figure}

Following an idea from \citet{1999A&A...345..221P},
\citet{2002ApJ...579..446M} first introduced a two-component model for
Mira stars. It was a disk limb darkened by a close warm molecular
layer. This model did well to account for spectral measurements in the
K and L bands. However, we noted some problems when using this model
to fit our measurements. First, because this model is spherically
symmetric, it cannot account for closure phase measurements different
from 0 or 180 degrees. Secondly, the physical parameters (temperature
and optical depth) present a degeneracy which cannot be resolved
without spectral information. We therefore decided to use a purely
geometrical description of the brightness distribution. This model
contains seven parameters describing the presence of a photosphere
with a center-to-limb darkening, a circumstellar envelope, and a spot
on the stellar surface (Fig.~\ref{fig:model}).

The photosphere is modeled by a limb-darkened disk. The center-to-limb
variation (CLV) is a simple, single-parameter, power law:
$I(\mu)/I(1)=\mu^\alpha$
\citep{1921ApJ....53..249M,1997A&A...327..199H}; with $\mu =
\cos(\theta)$, and $\theta$ the angle between the line of sight and
the radial vector from the center of the star. Compared to a more
classical 4-parameter law \citep[as in][]{2000A&A...363.1081C}, this
law has the advantage of fitting various CLV shapes with a single
parameter.  The circumstellar envelope is modeled by an annular ring
around the photosphere.  This is a simple way to account for the
existence of either (i) a warm molecular layer as proposed by
\citet{2004A&A...426..279P} or (ii) a molecular extension of the
photosphere as modeled by \citet{2008MNRAS.391.1994I}. Finally, we
needed a way to account for the asymmetry. We used the simplest model
available to explain an asymmetry: a single point-like spot. The model
allows either a ``cool'' (dark) or ``hot'' (bright) spot. There is no
restriction on the position of the spot (it is allowed to be outside
the photosphere).

To be sure that each one of these parameters are well constrained, it
is crucial to know the influence they have on the visibility
curve. The bottom panels of Fig.~\ref{fig:UDUG} give rough estimations
of the zone of influence of each parameter. At first order, the
molecular layer is constrained by the low frequencies, the disk size
by the first zero, the limb darkening by the height of the second lobe
and the asymmetry by the closure phase. Of course, to a lesser extent,
all the parameters affect each other at various degrees.

\subsection{Fitting the data} \label{sc:cp}

An advantage of this model is that it consists in a sum of brightness
distributions for which an analytical formula of the visibility
function exists. The function of a power law limb-darkened disk, of
parameter $\alpha$ and diameter $\theta_{\star}$, writes:
\begin{equation}
V_{\star}(v_r) = \sum_{k \geq 0} 
\frac{\Gamma(\alpha/2+2)}{\Gamma(\alpha/2+k+2)
  \Gamma(k+1)} \left( \frac{- (\pi v_r \theta_{\star})^2}{4} \right)^k\,,
\end{equation}
where $v_r$ is the radial spatial frequency ($v_r=\sqrt{u^2+v^2}$) and
$\Gamma$ the Euler function. The visibility function of an annular ring of diameter $\theta_{\rm layer}$ is:
\begin{equation}
V_{\rm layer}(v_r) = J_0(2 \pi \theta_{\rm layer} v_r)\,,
\end{equation}
where $J_0$ is the Bessel function of the first kind. Finally, the visibility function of a hot spot writes:
\begin{equation}
V_{\rm spot}(u,v) =  \exp\left(-2i\pi(X_{\rm spot}u+Y_{\rm spot}v)\right)\,,
\end{equation}
where the $X_{\rm spot}$ and $Y_{\rm spot}$ are the coordinates of the
spot (respectively right ascension and declination) relative to the
center of symmetry of the stellar surface.  The visibility of the full
model is the weighted sum of the three visibility functions, hence:
\begin{equation}
V(u,v) = F_{\star}V_{\star}(v_r)+F_{\rm layer}V_{\rm layer}(v_r)+F_{\rm spot}V_{\rm spot}(u,v)\,.
\end{equation}
$F_{\star}$,$F_{\rm layer}$ and $F_{\rm spot}$ are the relative fluxes
of, respectively, the disk, the envelope and the spot
($F_{\star}+F_{\rm layer}+F_{\rm spot}=1$). 

Closure phases are obtained by taking the argument of the product of
three complex visibilities: $\arg \left(V(u_1,v_1) \cdot V(u_2,v_2)
\cdot V(-u_1-u_2,-v_1-v_2)\right)$. Squared visibilities are derived
from the squared of the visibilities ($|V(u,v)|^2$). Because $u$ and
$v$ are wavelength dependent, accounting for the bandwidth smearing
required to average the squared visibilities before fitting them to the
data.  For the 3 first observation runs, they are averaged over the
1.59-1.69$\,\mu$m bandpass. For the last run, the model was
averaged over each channels bandpass ($\Delta \lambda = 40\,$nm). The
model is supposed to be achromatic.

Thanks to the analytical expressions, the model can be swiftly fitted
over a large range of parameter values. However, because the CP
measurements are at the same time very precise ($\approx 1$ degree)
and very sparse (there is only one phase measurements for three
squared visibilities), the spot position can have multiple $\chi^2$
minima. To find the most likely set of parameters, we used the
following strategy to achieve the \emph{global} optimization of the
$\chi^2$: for a grid of given spot positions, we first map the
$\chi^2$ minimized with respect to the other parameters. Then we use
the position which yields the best $\chi^2$ to initiate a local
optimization with respect to all parameters by a Levenberg-Marquardt
algorithm.

\begin{figure}
   \centering
\includegraphics[width=7.5cm]{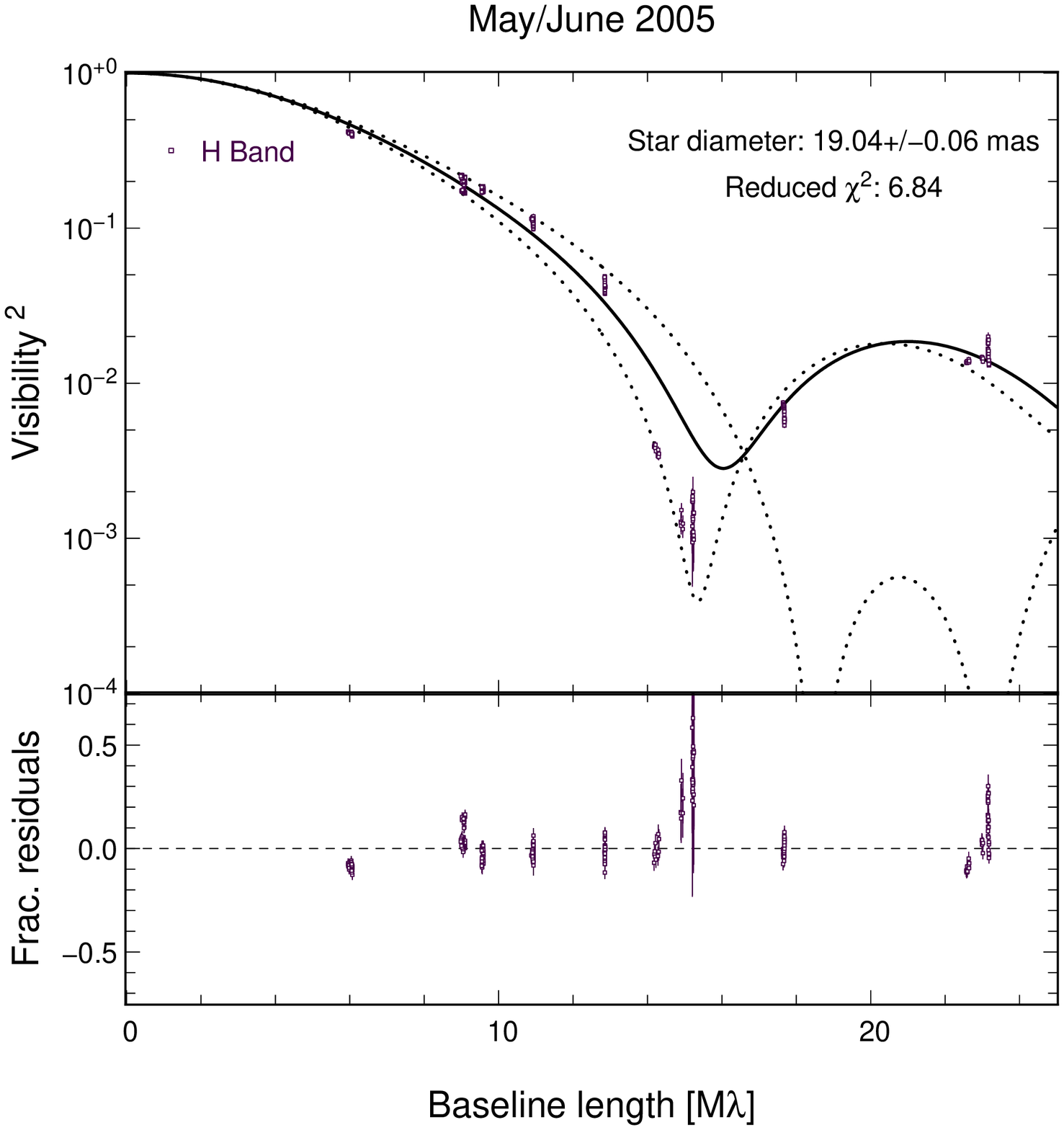} 
\includegraphics[width=7.5cm]{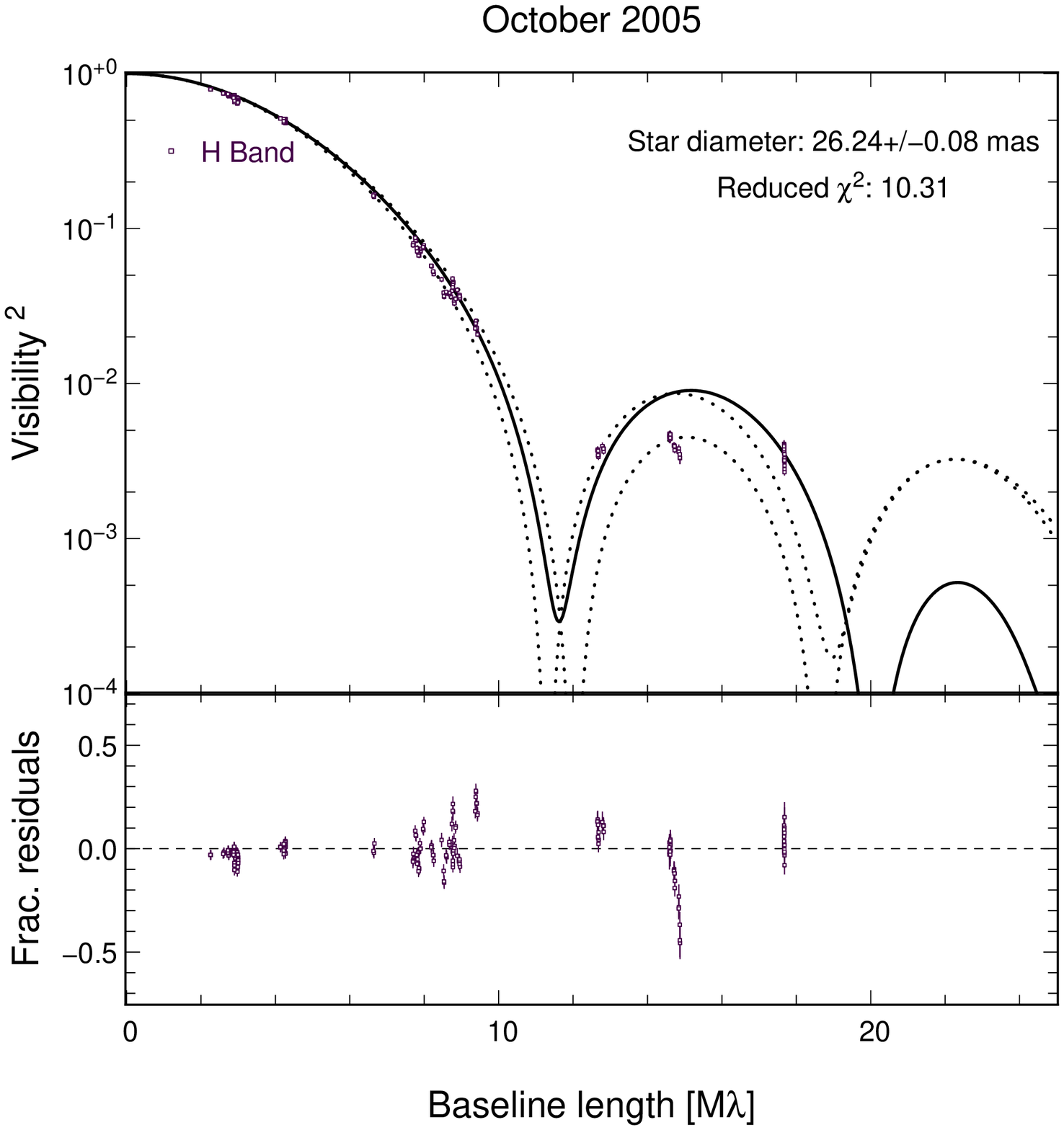} \\ \vspace{.1cm}
\includegraphics[width=7.5cm]{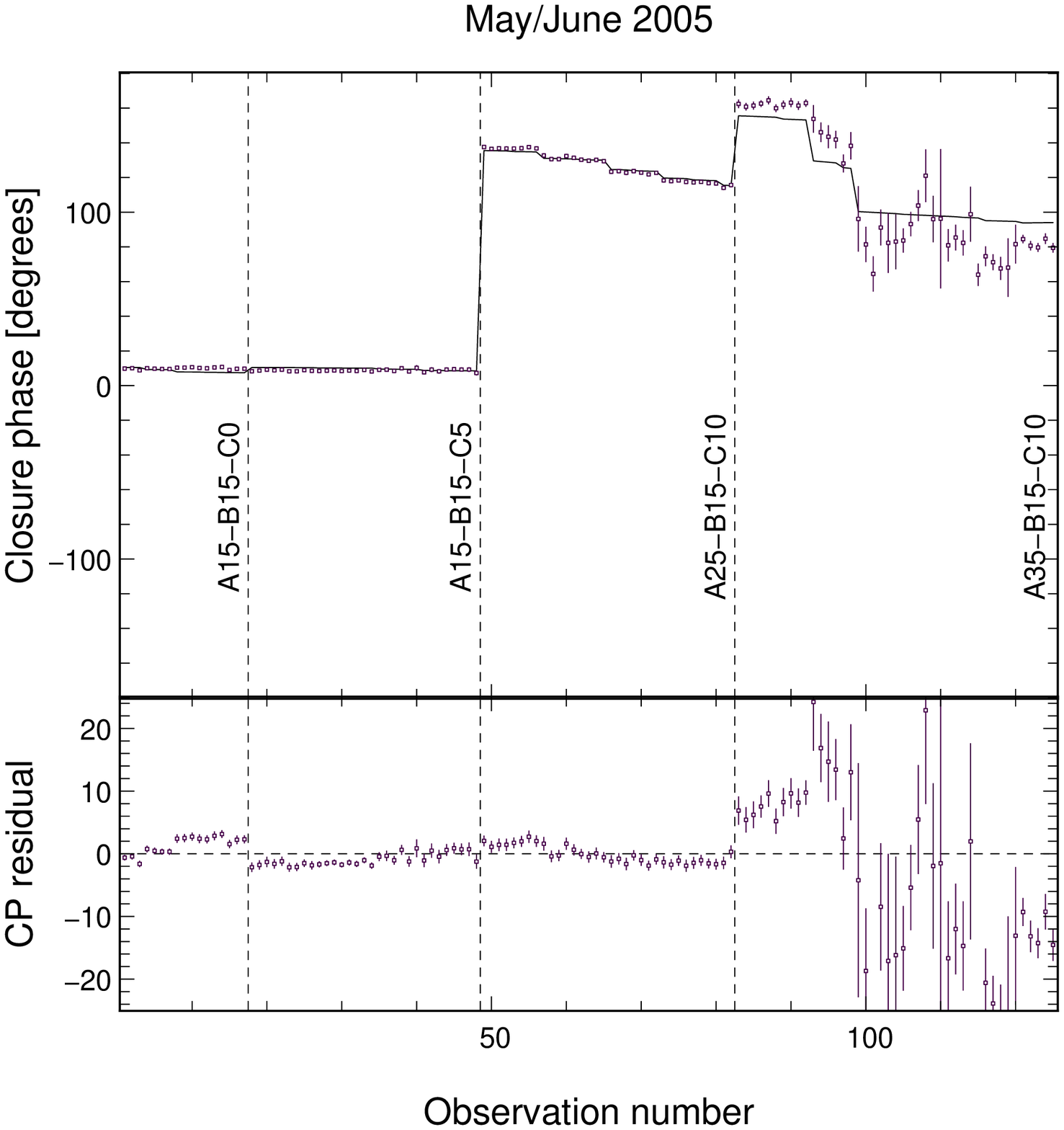}
\includegraphics[width=7.5cm]{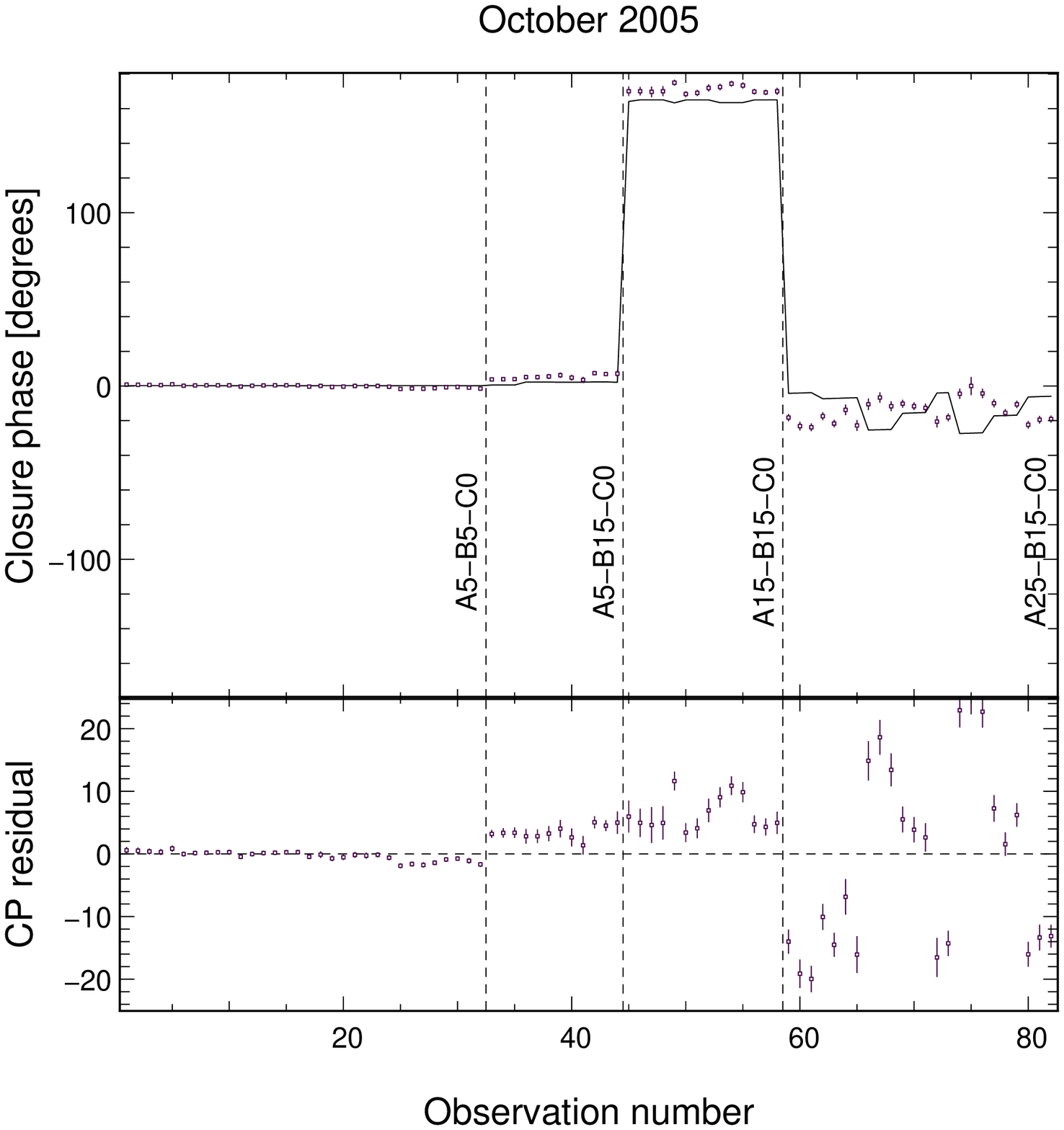}
      \caption{ Squared visibilities (upper panels) and closure phases (lower
        panels) at the two first epochs of observation. The curves
        correspond to the best fit of the model described in
        Fig.~\ref{fig:model}.  Three curves are plotted on the
        visibilities: the dashed curves are in the
        direction of the spot, and at 90 degrees from it. The solid
        curve is the visibility curve toward the longest baseline
        measured. The lower sub-panels show the residual errors.}
         \label{fig:Res_Vis1}
   \end{figure}

\begin{figure}
   \centering
\includegraphics[width=7.5cm]{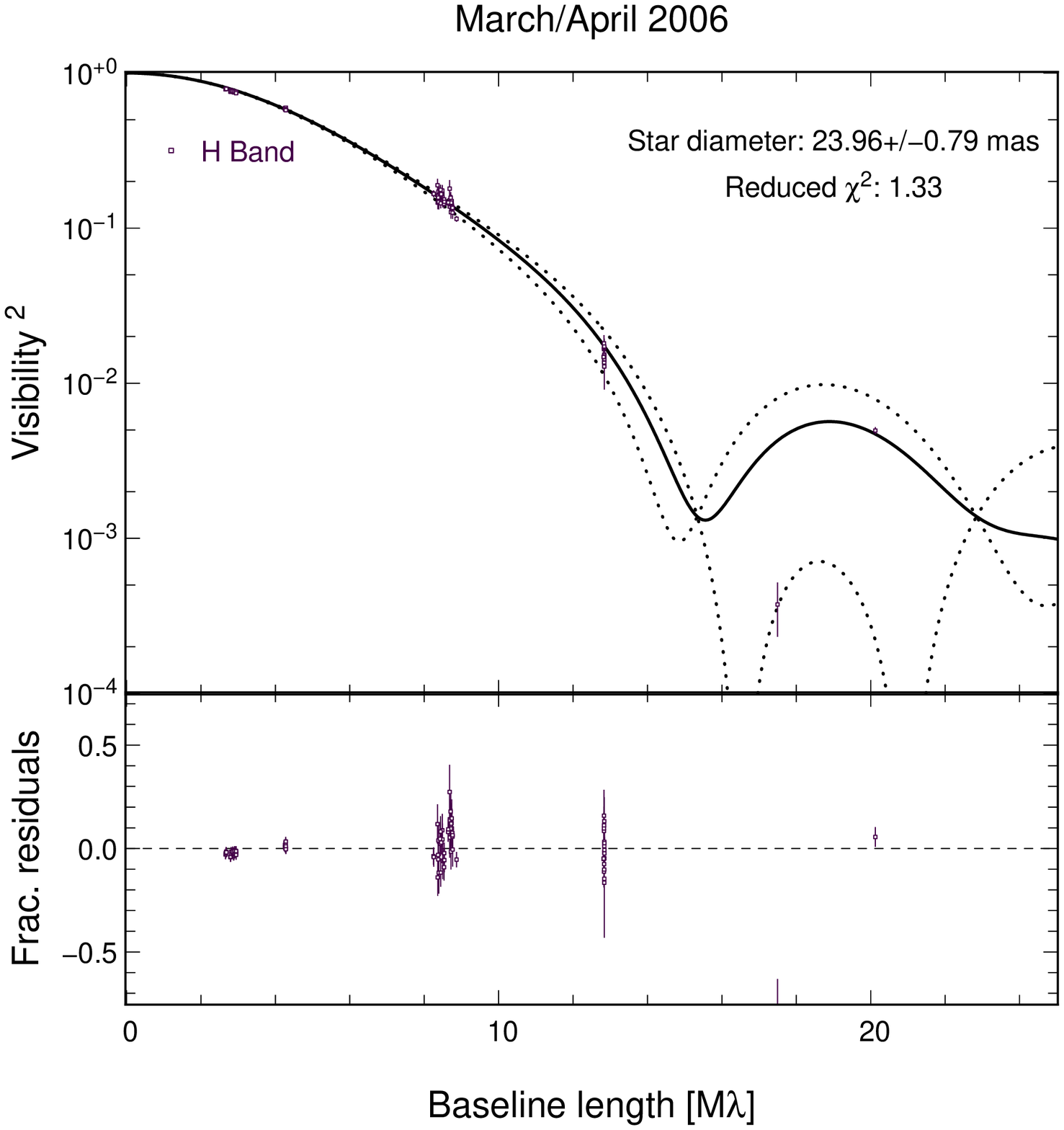} 
\includegraphics[width=7.5cm]{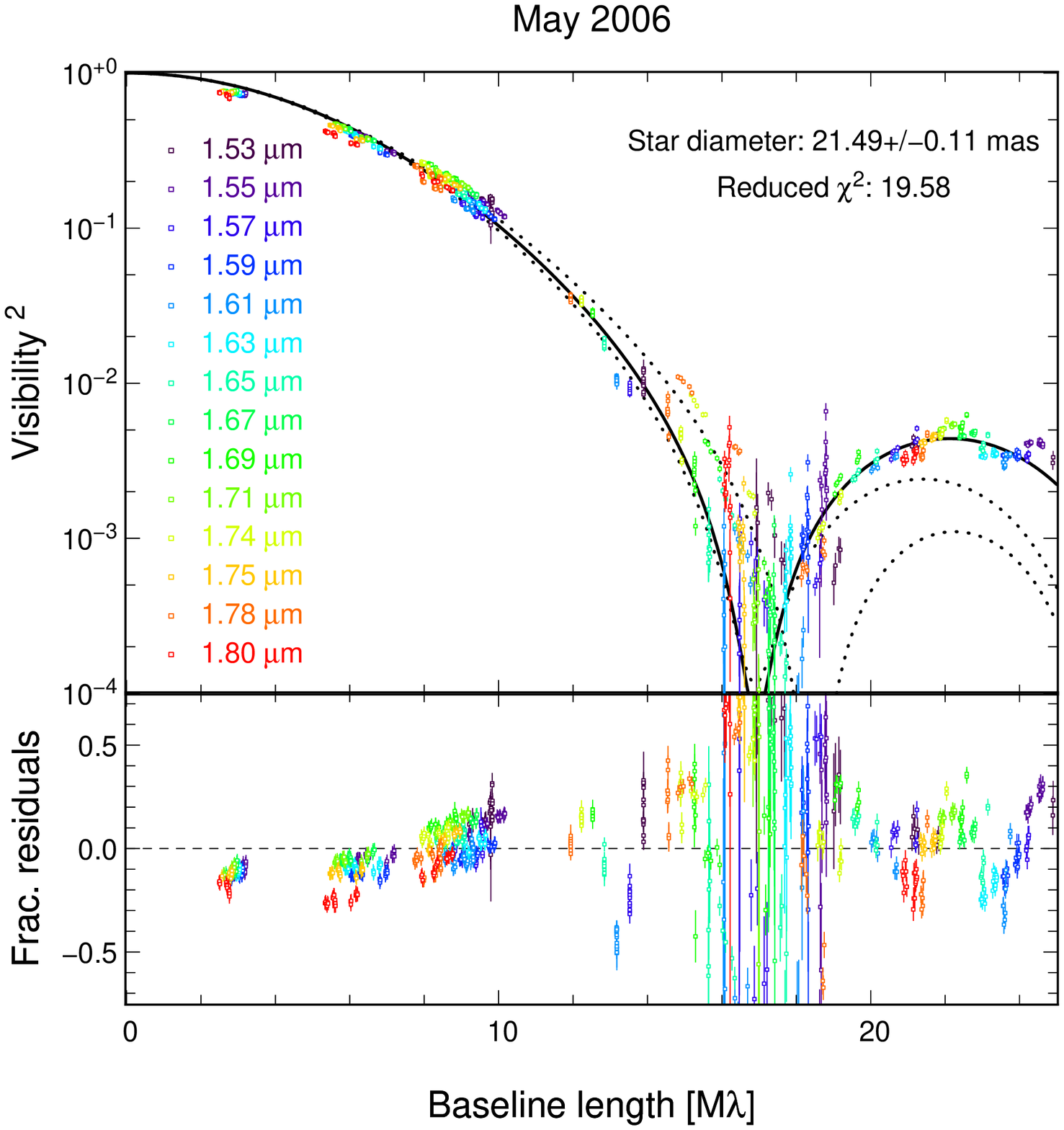} \\ \vspace{.1cm}
\includegraphics[width=7.5cm]{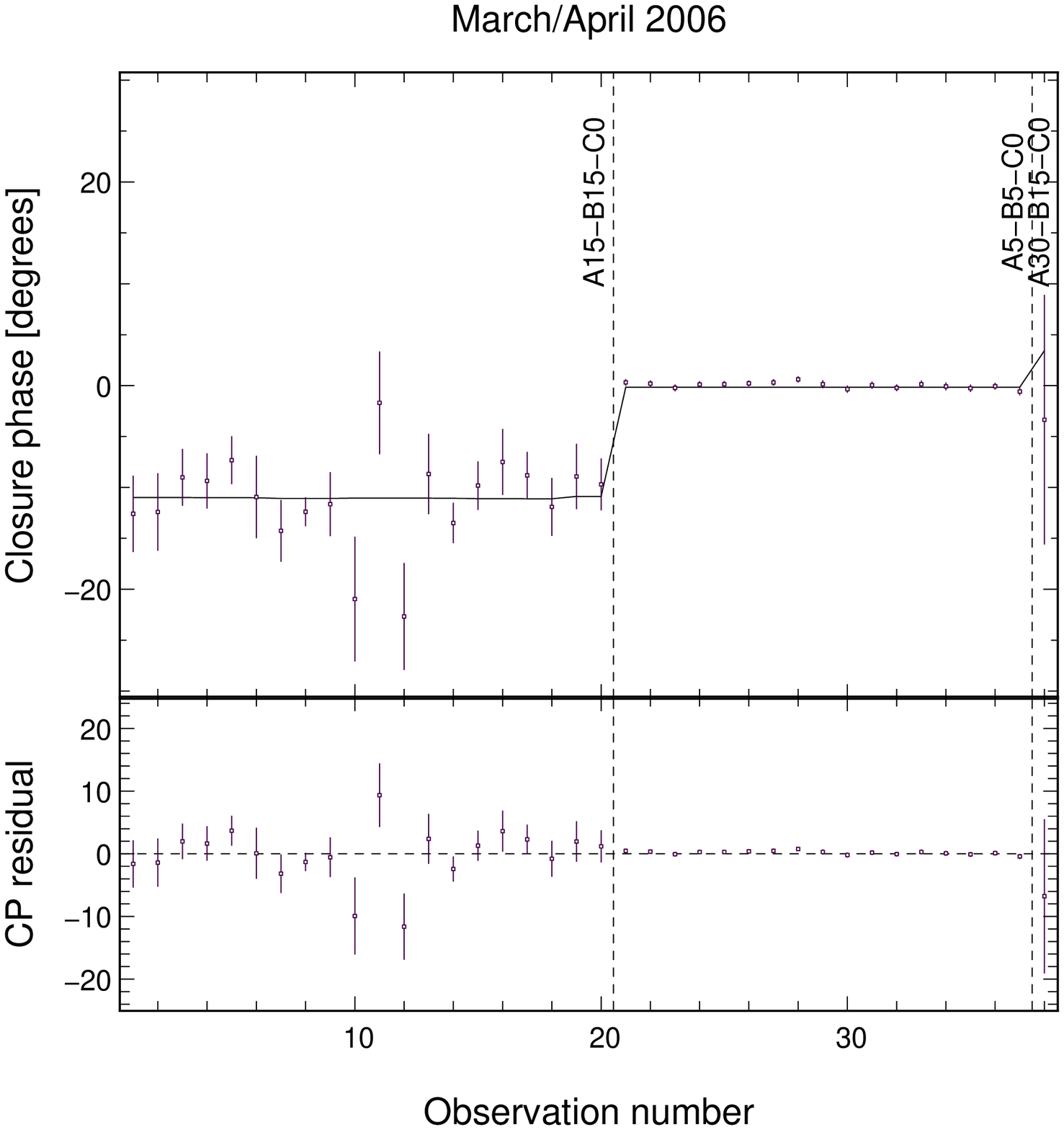}
\includegraphics[width=7.5cm]{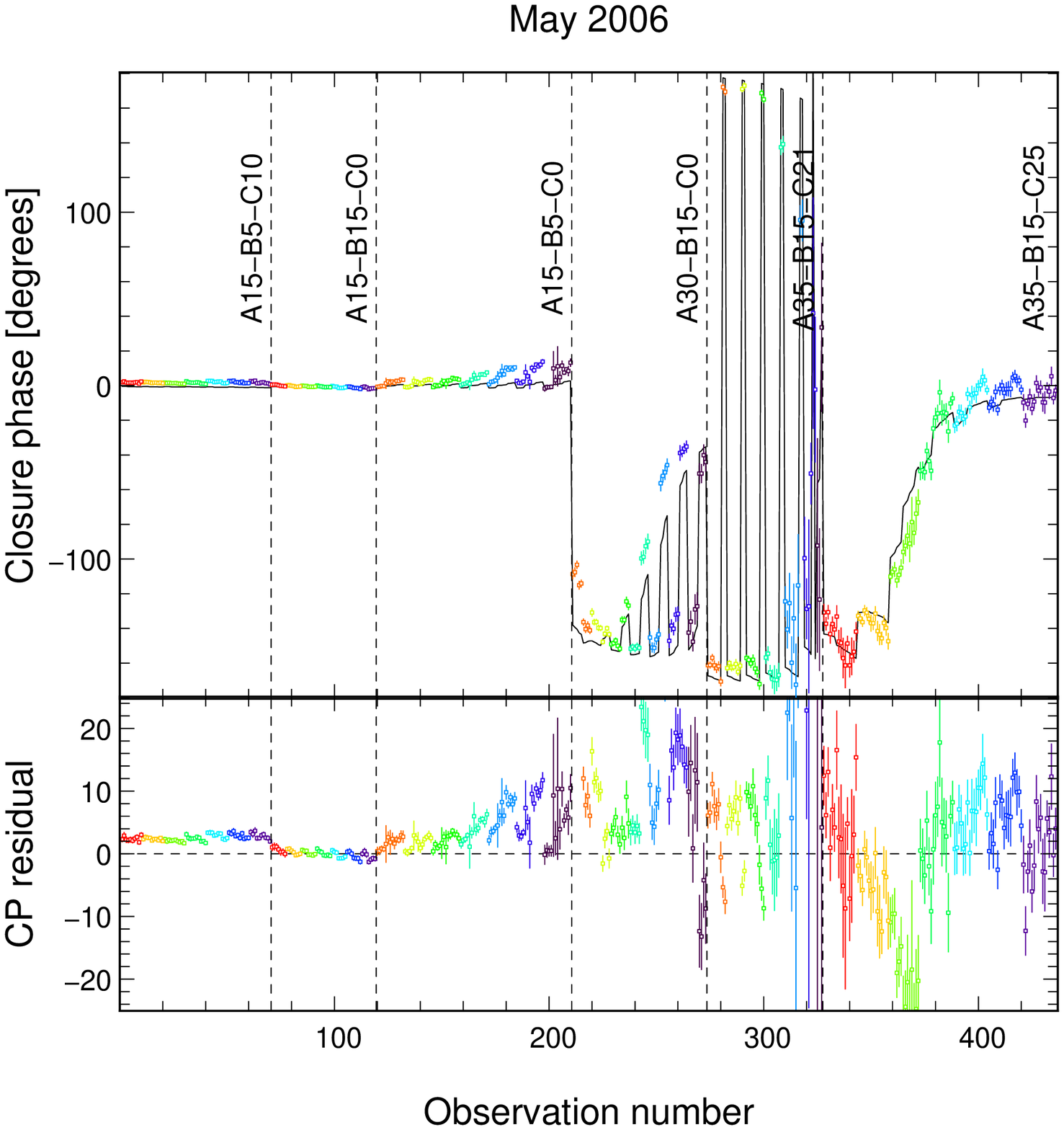}
      \caption{ Squared visibilities (upper panels) and closure phases (lower
        panels) at the two last epochs of observation. The curves
        correspond to the best fit of the model described in
        Fig.~\ref{fig:model}.  Three curves are present: the dashed
        visibility curves are in the direction of the spot, and at 90
        degrees from it. The solid curve is the visibility curve
        toward the longest baseline measured. The lower sub-panels
        show the residual errors.}
         \label{fig:Res_Vis2}
   \end{figure}

The upper panels of Fig.~\ref{fig:Res_Vis1} and~\ref{fig:Res_Vis2}
show the best fit of the model plotted on the squared visibilities (with
a logarithmic scale). Since the asymmetry cannot be represented by a
single radial visibility curve, we plotted three curves. The two
dashed visibility curves are in the direction and at 90 degrees of the
direction of the spot. The solid curve is the visibility curve toward
the longest baseline measured. The difference between the model and
the data points is plotted in the lower sub-panels. Residual errors on
the $V^2$ average around 1\%.

The lower panels of Fig.~\ref{fig:Res_Vis1} and~\ref{fig:Res_Vis2}
show the best fit of the model plotted on the closure phases. Even
though the point-like spot model is a very rough estimation of what
the asymmetry could be, the general agreement between the model and
the data confirms the validity of using such a simple representation.
However, the results cannot exclude more complicated asymmetries, like
the presence of multiple spots or other heterogeneities. For the
October observations for example, small variations in the CP are not
well matched by our model, hinting for a more complex repartition of
the asymmetry.

We noted that the minimum reduced $\chi^2$ can be quite different from
1, ranging from a value of 1.3 (March 2006) to 19.6 (May 2006). Lower
$\chi^2$ can be obtained by using more complex models, like fitting
two spots instead of one, or adjusting a chromatic limb darkening to
the May 2006 data. However, the multiple parameters in such a case
were too badly constrained to allow good determination. In the end, we
decided to stay with our most simple model which gives a good
compromise between fitting the data well and a reasonable number of
free parameters.


\subsection{Parametric imaging} 

\begin{figure}
   \centering
   \includegraphics[width=15cm]{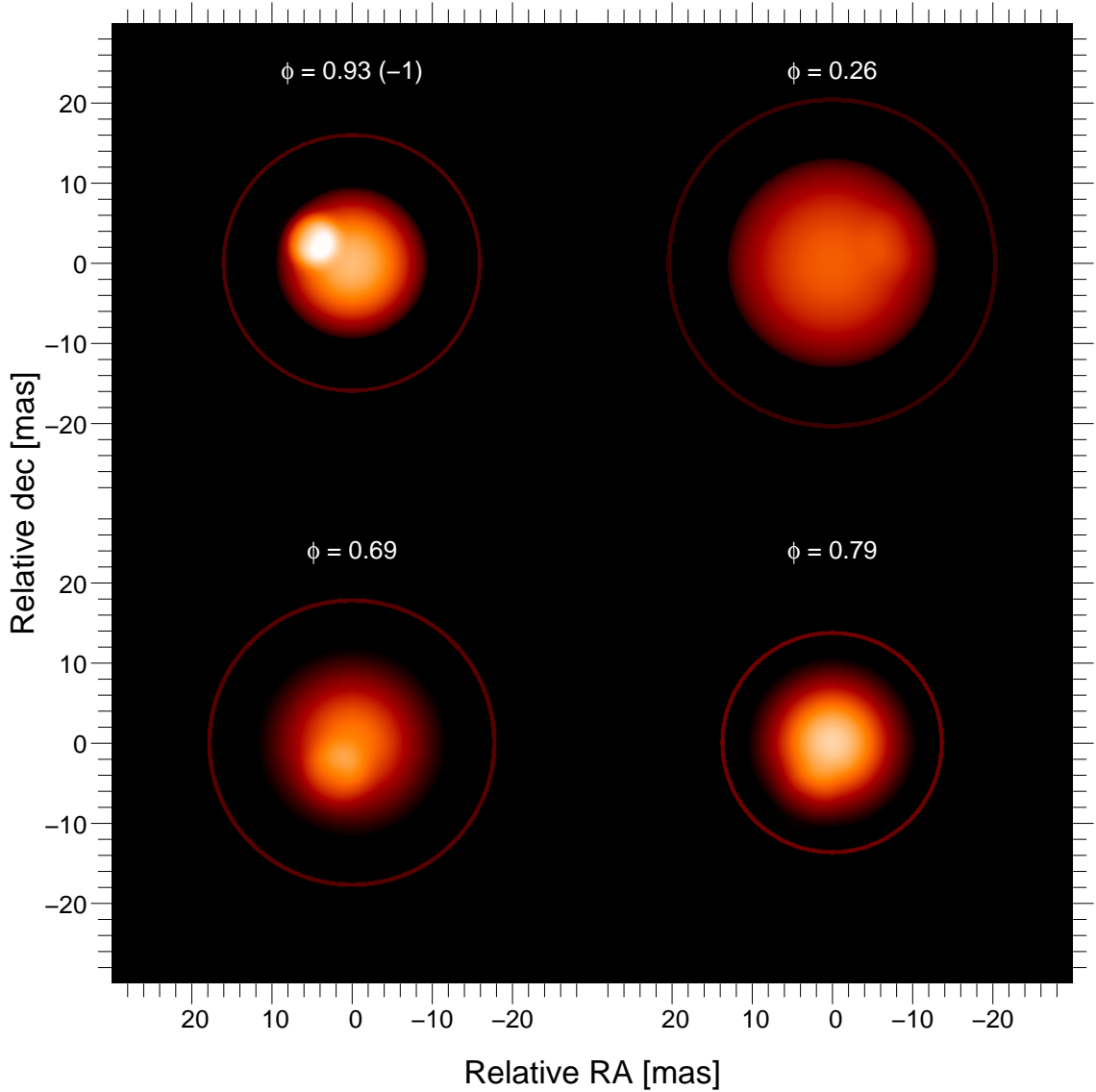}
      \caption{Parametric imaging. The parametric model is showed in
        Fig.~\ref{fig:model}. The values used are the ones reported in
        Table~\ref{tb:Res_fit}. Apparency of the spot was chosen to be
        a fifth of the stellar diameter, below the resolution power of
        the interferometer.  The relative brightnesses are normalized
        according to H band magnitude from Table~\ref{tb:flux}.  At $\chi$
        Cygni's distance of 170\,pc, 5.9\,mas correspond to 1\,AU
        (Sect.~\ref{sc:para}).  }
         \label{fig:Param_Image}
   \end{figure}

The results and error bars of the fits can be found in
Table~\ref{tb:Res_fit}. From these values, we can retrieve a
brightness distribution of the object, ie, an image. The resulting
images are presented in Fig.~\ref{fig:Param_Image}, and should be
compared to the regularized images of
Fig.~\ref{fig:Reg_Image}. Concordance is quite convincing, with a
remarkable reproductivity in terms of diameter and position of the hot
spots. The main difference is the presence of the molecular layer,
situated around 0.5 stellar radius above the photosphere. The
faintness of the layer (a faintness expected within our wavelength
range of observation) may explain why the regularization algorithm did
not image it. This on top of the fact that the regularization term
does favor an empty environment.

\section{Linking spatial to physical parameters} \label{sec:Physics}

\subsection{Stellar diameter} 
\subsubsection{Rosseland radius}

In the case of an extended-atmosphere star such as Mira variables, a
sensible definition of the radius as to be agreed upon. A quantity
mostly used in Mira modeling is the Rosseland radius. This is an
optical depth radius ($R_{\rm Ross} = r$ where $\tau_{\rm Ross} = 1$)
and unfortunately not an observable quantity. In the case of the solar
disk, the photospheric radius of the Sun is defined by the position of
the CLV inflection point.

The diameter values stated in Table~\ref{tb:Res_fit} correspond to the
furthest emission point of the photosphere, molecular layer
notwithstanding. The problem of this definition of diameter is that it
is highly model dependent.  Such influence is emphasized in
Table~\ref{tb:fit}. The table compares the results of three different
models fitted to the dataset of two epochs. It shows how profound the
difference on diameter measurements can be depending on the model,
well outside the range of error bars.  A solution could have been to
consider only the uniform disk radius, and to use tables \citep[for
  example in][]{2000MNRAS.318..387D} to derive the Rosseland radius
from simulations
\citep{2004MNRAS.355..444I,2004MNRAS.352..318I}. However, this
technique poses the problem of fitting a uniform disk on a dataset
which is not compatible with it: the bias on the UD diameters depends
on the Fourier coverage \citep{2009arXiv0904.2166P}.

However, a noticeable advantage of this work is the good coverage of
spatial frequencies at all epochs. Our model, which covers a large
range of CLV possibilities, allows to disentangle the main features of
the stars (layer, spot). Simulations of pure-continuum brightness
profiles have steep flanks which mark the position of
continuum-forming layers \citep{1987A&A...186..200S}. We are confident
that the steep flank observed on the fitted model marks, as rightly as
possible, the limit of the Rosseland Radius.

\subsubsection{Comparison with other interferometric observations}

Numerous $\chi$ Cyg diameter measurements are present in the
literature, but the multiple techniques of determination makes
comparison difficult. The first near-infrared interferometric
observations of $\chi$ Cyg were obtained by
\citet{2000MNRAS.318..381Y} using the COAST instrument. In the 1.3
$\mu$m continuum bandpass, they obtained a Gaussian FWHM of
$13.9\pm0.8$ mas at $\phi = 0.83$. The 44\% discrepancy with our
diameter of $19.04 \pm 0.09$ mas at $\phi = 0.93$ is difficult to
explain by the difference of phase only. However it can be explained
by the study of \citet{1998A&A...339..846H}. In their paper, they
showed that the ratio between continuum radius and Gauss radius should
be around 0.6, a value close to what is observed here.
\citet{2004A&A...426..279P} used the beam combiner FLUOR on IOTA to
obtain narrow band measurements in the K-band. They fitted a more
complete model including a uniform disk and a molecular layer with a
wavelength dependent optical depth. They found an apparent stellar
diameter of $21.10 \pm 0.02$ mas at $\phi = 0.24$. This value is
compatible with our measurements of $20.90 \pm 0.12$ mas obtained by
fitting a model without limb darkening, but not with the diameter of
$26.25\pm0.08$ mas measured when accounting for the CLV shape. So the
discrepancy is logically explained by the fact that they did not
account for the limb-darkening.

Another important diameter value still to be mentioned is the one
obtained by \citet{2003ApJ...589..976W} using the ISI heterodyne
interferometer. They reported a diameter value of $39.38\pm4.02$ mas
at $\phi = 0.51$ in the 11 $\mu$m continuum bandpass. Even though the
phase of observation is different from ours, this value is far from
being compatible. A possible interpretation could be the one proposed
by \citet{2004A&A...424.1011O}, suggesting the presence of a warm,
spectroscopically ``hidden'', H$_2$O molecular layer. It is
interesting to note that the 11 $\mu$m stellar diameter measurements
match our measurements of the position of the molecular layer,
something already noticed by \citet{2004A&A...426..279P}.

Concerning the limb darkening, this paper offers the first
measurements on a Mira star.  But other observations exist on other
types of stars.  \citet{2006A&A...453..155M} reported good fits with
$\alpha = 0.16$ toward Polaris and $\delta$ Cepheid, two variable
Cepheids. \citet{2006A&A...460..855W} and \citet{2008A&A...485..561L}
observed M giant stars and obtained respectively $\alpha =0.24\pm0.03$
for Menkar and $\alpha =0.258\pm0.003$ for Arcturus. In comparison,
$\chi$ Cygni's $\alpha$ values (greater than 1) look large. However,
one has to consider the size and pulsation of the photosphere of Mira
variables. In that respect, our values are compatible with the CLV
simulations presented in \citet{2002MNRAS.336.1377J}.

\subsubsection{Effective temperature} \label{sc:eff_temp}

The effective temperature of a star depends upon its angular diameter
and its bolometric flux,
\begin{equation}
\sigma \,\cdot\, T_\star^4 = \frac{4}{\theta_\star^2} \,\cdot\, F_{\rm Bol} \,
\label{eq:tp}
\end{equation}
where $\sigma$ is the Stefan-Boltzmann radiation constant, and
$F_{\rm Bol}$ is the observed flux, integrated over all
wavelengths. \citet{2000MNRAS.319..728W} reported J, H, K, and L band
observations of the star at several phases. Thus, a phase-dependent
bolometric flux can be estimated by fitting a black body distribution
on the reported magnitudes. In Table~\ref{tb:flux} we present the
photometric estimations as well as the bolometric flux at the epochs
of observations. Applying Eq.~\ref{eq:tp}, we deduced the effective
temperatures (reported in Table~\ref{tb:physics}). They are
interesting in the sense that they are much cooler than what was
deduced in \citet{2004A&A...426..279P} and are consistent with an M8
spectral type \citep{1998A&A...331..619P}.

\subsubsection{A time-variable view of $\chi$ Cygni} \label{sc:var_temp}

\begin{figure}
\centering
\includegraphics[width=13cm]{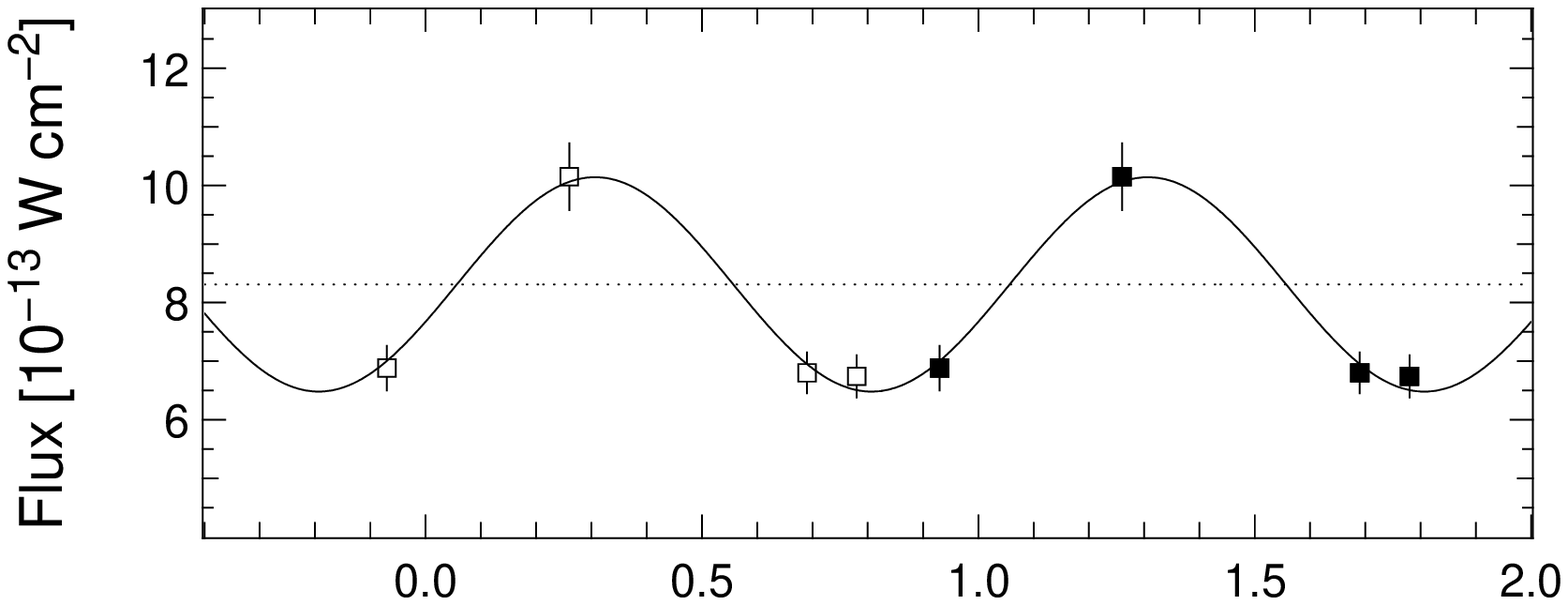} \\
\includegraphics[width=13cm]{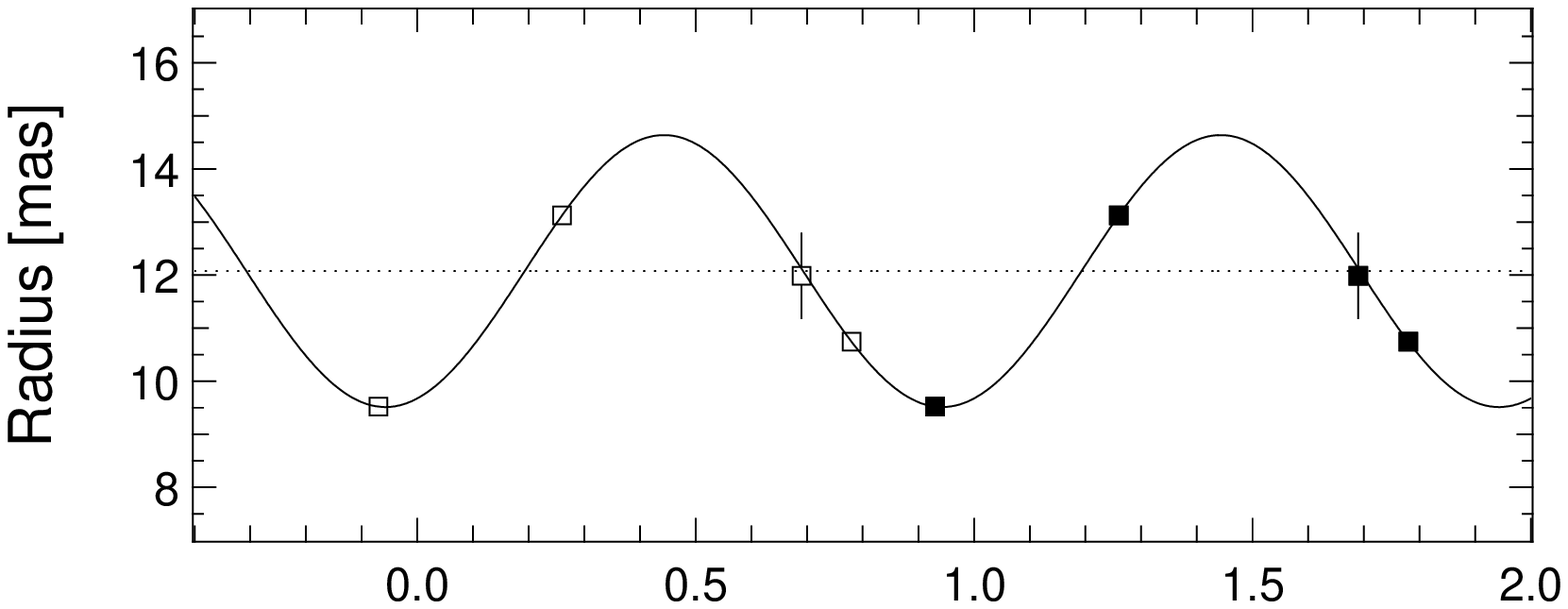} \\
\includegraphics[width=13cm]{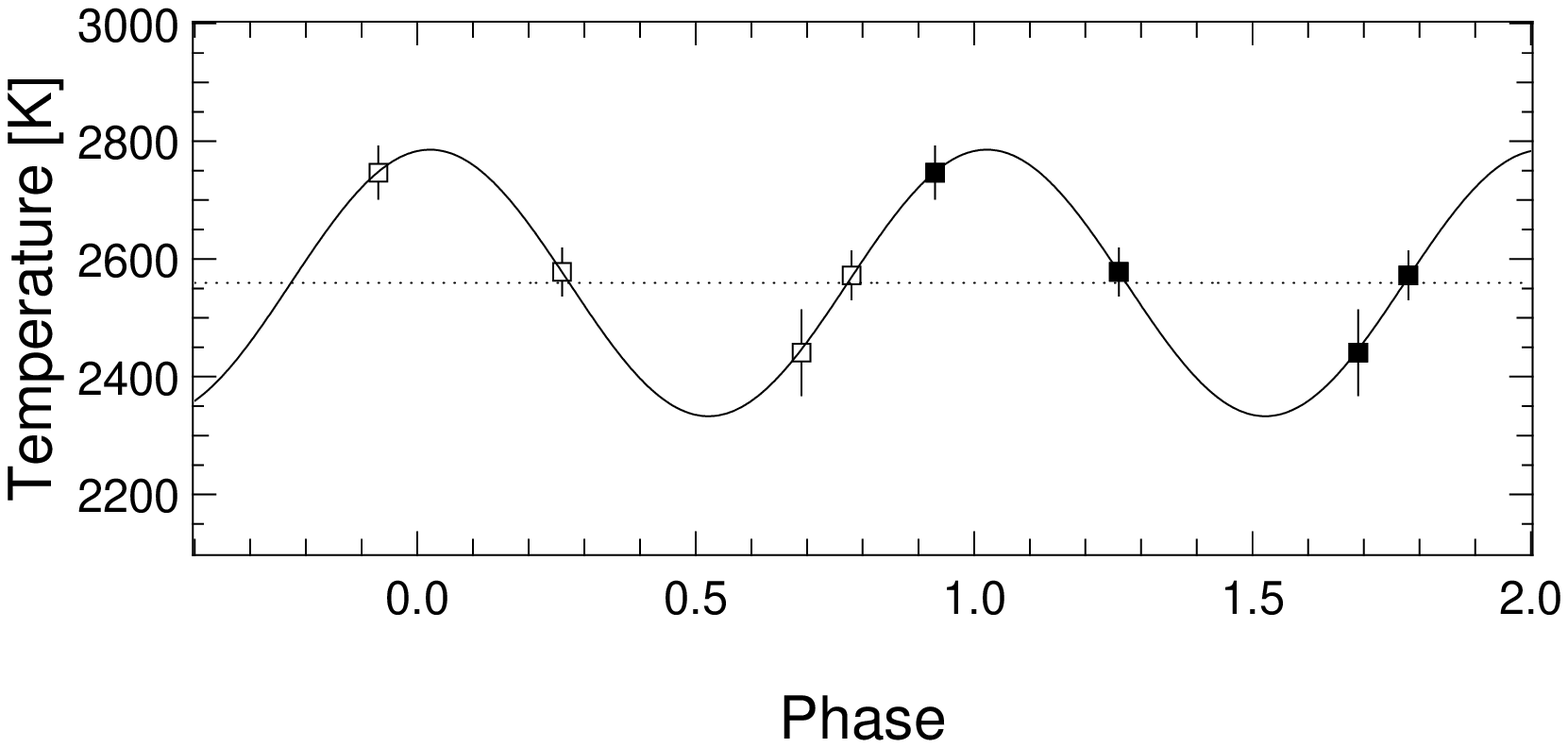}
\caption{Temporal evolution of bolometric flux, radius and temperature
  of the photosphere of $\chi$ Cyg. Values (Table~\ref{tb:physics})
  are plotted twice with $+1$ phase shift (white and black dots). A
  sinusoidal is fitted to the data points, giving a linear radius of
  12.1\,mas and an average effective temperature of 2560\,K
  (Table~\ref{tb:evolution}). The temperature was derived according
  to: $T_{\rm eff}=(4 \sigma F_{\rm Bol}/\theta_\star^2)^{1/4}$. It
  follows a periodic variation phase-shifted by 0.58 compared to the
  radius, ie, almost opposed to the dilatation of the photosphere. }
\label{fig:Temp_evol}
\end{figure}

Fig.~\ref{fig:Temp_evol} represents the flux, radius and temperature
as a function of time. We fitted a sinusoidal model to the data. Best
fit values are reported in Table~\ref{tb:evolution}.  Linear radius is
equal to 12.1\,mas and the amplitude of the pulsation is 5.2\,mas
(43\%). Minimum diameter happens at $\phi = 0.94\pm 0.01$. As
expected, the temperature $T$ is anti-correlated with the diameter,
with a slight lag of 8\% of the period ($\approx 30$\,days). This
anti-correlation, as well as a similar lag, is also observed on
Cepheids stars \citep{2005AJ....130.1880A}.

On the other hand, the bolometric flux is mostly correlated with the
diameter, an indication that the variation in the bolometric flux is
dominated by the variation of the size of the object. The temperature
plays a minor role on the bolometric flux, but making it phase shifted
by 0.14 in advance of phase compared to the radius. It is interesting
to note that the visual magnitude, strongly affected by the molecular
environment, is not in phase with the bolometric flux, but instead is
in phase with the temperature.

There are not many time-variable radius and temperature figures in the
literature. \citet{2000MNRAS.318..381Y} observed a variation with a
maximum diameter at phase 0.6, but observations were done at a shorter
wavelength (905\,nm) which is affected by TiO absorption/emission. No
noticeable diameter variation was observed in the less affected
1.3\,$\mu$m bandpass. On the other hand, \citet{2002ApJ...570..373T}
observed a clear variation in diameter in the near infrared (K band)
toward the Mira star R\ Tri Mira. They observed a 10\% decrease in the
UD diameter between phase 0.77 and 0.91. It agrees with the variation
in radius observed here on $\chi$ Cyg, but unfortunately the time
coverage on R\ Tri does not allow a more through-full
comparison. Noteworthy is the work done by
\citet{2008ApJ...673..418W}: they derived time-dependant UD diameter
for eight Mira variable, including $\chi$ Cygni. Minimum diameter is
observed at phase $\approx 0.7$, a value inconsistent with our
work. This could be explained by the fact that they fitted UD
diameters which do not account for the molecular environment of the
star.

Least, a plot similar to the Fig.~\ref{fig:Temp_evol} is present in
the Ph.D. Thesis of \citet{1973.Strecker}. He used bolometric flux and
3.5$\mu$m observations of $\chi$ Cyg to deduce a time-variable
temperature and stellar diameter. In terms of variations, his results
match our dataset well, with a minimum diameter around 0.9 and a
minimum temperature at $\phi =0.5$. However, compared to our work, he
overestimated the size of the star by a factor 1.5 ($\approx36\,$mas)
and underestimated its effective temperature ($\approx 2000\,$K).

\subsection{Temperature and opacity of the molecular layer}  \label{sc:opa_l}

The highly chromatic brightness of the layer observed by
\citet{2004A&A...426..279P} immediately suggested the presence of
molecules, mainly H$_2$O and CO. Because of its molecular nature, it
is extremely difficult to disentangle the relative effect of
temperature and opacity on the brightness.

To derive a first order estimation of the temperature, we used the
assumption of a grey atmosphere. Hence, from \citet{1997ApJ...476..327R},
the temperature writes:
\begin{equation}
  T_{\rm
  layer}^4=T_\star^4 \left(
  1 - \sqrt{1-(\theta_\star/\theta_{\rm layer})^2} \right) \,.
\label{eq:gray}
\end{equation}

From this first order estimation of the temperature, we can also
deduce the optical depth of the molecular layer ($\tau_{\rm layer}$)
by using the flux conservation relation:
\begin{equation}
\frac{F_{\rm layer}}{F_\star}=\frac{B(\lambda,T_{\rm layer})}{B(\lambda,T_\star)}\,\cdot\,\frac{\theta_{\rm layer}^2}{\theta_\star^2}\,\cdot\,\frac{1-\exp{(-\tau_{\rm layer}})}{\exp{(-\tau_{\rm layer})}}\,.
\end{equation}
where $B(\lambda,T)$ is the Planck function and $\lambda$ the
wavelength. Hence:
\begin{equation}
\tau_{\rm layer} = \ln{\left(1+\frac{F_{\rm layer}}{F_\star} \,\cdot\, \frac{
    B(\lambda,T_\star) \, \theta_\star^2}{ B(\lambda,T_{\rm
    layer}) \, \theta_{\rm layer}^2} \right)} \,.
\label{eq:tau}
\end{equation}
The temperature and optical depth are reported in
Table~\ref{tb:physics}. With an effective temperature ranging from
1750\,K to 2000\,K, the layer is significantly cooler than the
excitation state of CO ($\Delta v =3$) molecules as observed by
\citet{1982ApJ...252..697H}. On the other hand, the temperature agrees
well with multi-wavelength interferometric observation from
\citet{2004A&A...426..279P}. The optical depth of the molecular layer
is very low ($\tau <0.1$), something we expected in the H band
\citep[and already reported on other Mira stars, as toward S\ Ori
  in][]{2008A&A...479L..21W}.

\section{Deriving Mass and Distance from the kinematics of the atmosphere} \label{sc:der_ma}

\subsection{On the existence of a molecular layer}
 
There is a strong debate about the physical nature of the warm
molecular layer as seen by optical interferometry. Precise
interferometric values \citep{2004A&A...426..279P} and image
reconstruction work \citep{2009A&A...496L...1L} tend to see a
shell-like structure. Simulations \citep{2003SPIE.4838..163S} would
favor a continuum emission from the photosphere up to a certain height
in the atmosphere. This work does not pretend nor wish to resolve this
issue. In the previous section, we chose to use the shell-like layer
approach to understand our data, knowing we are lacking the resolving
power to convincingly distinguish between the two. From a geometrical
point of view, the only difference between the two approaches only
lies in the width of the layer, which is negligible in a shell-like
representation.

Whatever the width of the molecular layer, clues on its kinematics are
offered by dynamical modeling of the pulsating atmosphere
\citep{1985ApJ...299..167B,1988ApJ...329..299B,1996A&A...307..481B}.
Specifically, \citet{1985ApJ...299..167B} showed that the atmosphere
should be periodically criss-crossed by supersonic shocks. They
computed a post-shock density 59 times the density of the pre-shock,
forming a zone in which rapid cooling would allow high nucleation
rates \citep{2000ARA&A..38..573W}. The material is then inwardly
accelerated by the gravitational field, passing the sonic point
closely below the shock-front. At this time, pressure forces become
unimportant, making the rest of the trajectory ballistic-like. This
theory nicely explains how warm molecules -- and eventually grains --
could be formed so low within the atmosphere of the star.

If this model holds, the formation rates would decide the width of the
molecular layer: density and temperature would define a spatially
delimited zone in which molecules would happen to be more concentrated
than in other atmospheric areas.

\subsection{A modified Baade-Wesselink method to derive the distance}

\subsubsection{The CO ($\Delta v=3$) absorption features}

The \citet{1926AN....228..359B} and \citet{1946BAN....10...91W} method
consisted originally in deriving the absolute diameter and distance of
a star by means of photometry and spectroscopy. The interferometric
Baade-Wesselink method \citep[applied to Cepheids
  in][]{2004A&A...416..941K} differs by using a direct angular
diameter measurement instead of photometry. We propose here to adapt
this method to Mira stars, using the angular acceleration of the
molecular shell. Concretely, the method allows deriving the distance
by the relation:
\begin{equation}
{\rm parallax\ (mas)} = \cfrac{1}{p} \cdot \cfrac{g_{\rm angular}\ ({\rm mas/s}^{2})}{g_{\rm radial\ velocity}\ ({\rm AU/s}^{2})}\,,
\label{eq:para}
\end{equation}
where $p$ is a projection factor, $g_{\rm angular}$ the geometric
acceleration observed by interferometry, and $g_{\rm
  radial\ velocity}$ the absolute acceleration observed by radial
velocity. An alternative method could have been to use the speed
instead of the acceleration (first derivative of size instead of
second derivative). The advantage of using the acceleration is to
avoid a bias due to uncertainties in the local stellar velocity.

Radial velocity of the molecular shell can be obtained through the CO
second overtone vibration-rotation transitions ($\Delta v=3$),
observable by spectroscopy around 1.6 $\mu$m
\citep{1982ApJ...252..697H}. The high vibrational energy of these
molecules is characteristic of a warm environment (CO molecules in
lower vibrational states are also observed but probe a cooler
environment farther away from the photosphere).  These excited
molecules follow a roughly similar path on most of Mira stars, tightly
correlated with the visual phase: they are created around maximum
visual brightness, steadily accelerated toward the star, and destroyed
at the following stellar maximum
\citep{1978ApJ...220..210H,1984ApJS...56....1H,1995ASPC...83..399H,
  1997AJ....114.2686H,2005A&A...431..623L}. The excitation temperature
is also correlated with the phase, showing a neat exponential cooling
of the falling material. Line doubling is sometimes observed when the
rotationally hot molecules form before complete dissociation of the
less excited ones. This behavior is making considerable sense in the
light of a \citet{1985ApJ...299..167B} scenario summarized in the
previous section.

One should be aware that using the Baade-Wesselink method in these
conditions makes use of two strong hypotheses:
\begin{itemize}
\item The molecular layer expansion is radial.
\item The molecular layer as seen by interferometry is the one from
  which originates the CO ($\Delta v=3$) absorption.
\end{itemize}
The radial nature of the CO displacement is a likely assumption
because of the cycle to cycle repeatability as observed by
\citet{1982ApJ...252..697H}. The question of the concomitance of the
molecular layer with the CO ($\Delta v=3$) molecule is sensible,
especially with respect to the unknown width of the molecular
layer. Regarding the effective temperature, the depth of the
absorption lines reported by \citet{1982ApJ...252..697H} match
1700\,K, close to the temperature of the layer reported in
Table~\ref{tb:physics}. On the other hand, the excitation temperature
is higher, but the exponential decrease from 4400\,K to 2200\,K
observed by \citet{1982ApJ...252..697H} is hinting that excitation and
effective temperatures are unlikely to be in equilibrium. 

In the near future, we expect high spectral resolution
interferometers to lift this uncertainty by probing the apparent
angular diameter of the star directly within the absorption line.

\subsubsection{The projection factor} \label{sc:p}

The projection factor is one of the limiting parameters of the
Baade-Wesselink methods. However, in the case of a Mira star imaged by
interferometry, we can use the brightness distribution of the
photosphere and combine it with the photosphere and molecular layer
angular diameter. Assuming a shell of zero thickness, the projection
factor writes:
\begin{equation}
p=\cfrac{\int_0^{R_\star}2\pi r \mu_\star^\alpha dr}{\int_0^{R_\star}2\pi r \mu_\star^\alpha \mu_{\rm layer} dr}\,,
\label{eq:p}
\end{equation}
where $\mu_\star$ and $\mu_{\rm layer}$ are the projected radii, ie,
$\mu_\star=(1-(r/R_\star)^2)^{1/2}$ and $\mu_{\rm layer}=(1-(r/R_{\rm
  layer})^2)^{1/2}$. Using this formula and the numbers stated in
Table~\ref{tb:physics}, the projection factor at the four phases of
observations range from 1.22 to 1.27. In the following section we will
use the projection factor $p = 1.245\pm0.025$.

\subsubsection{$\chi$ Cygni parallax} \label{sc:para}

\begin{figure}
\centering
\includegraphics[width=13cm]{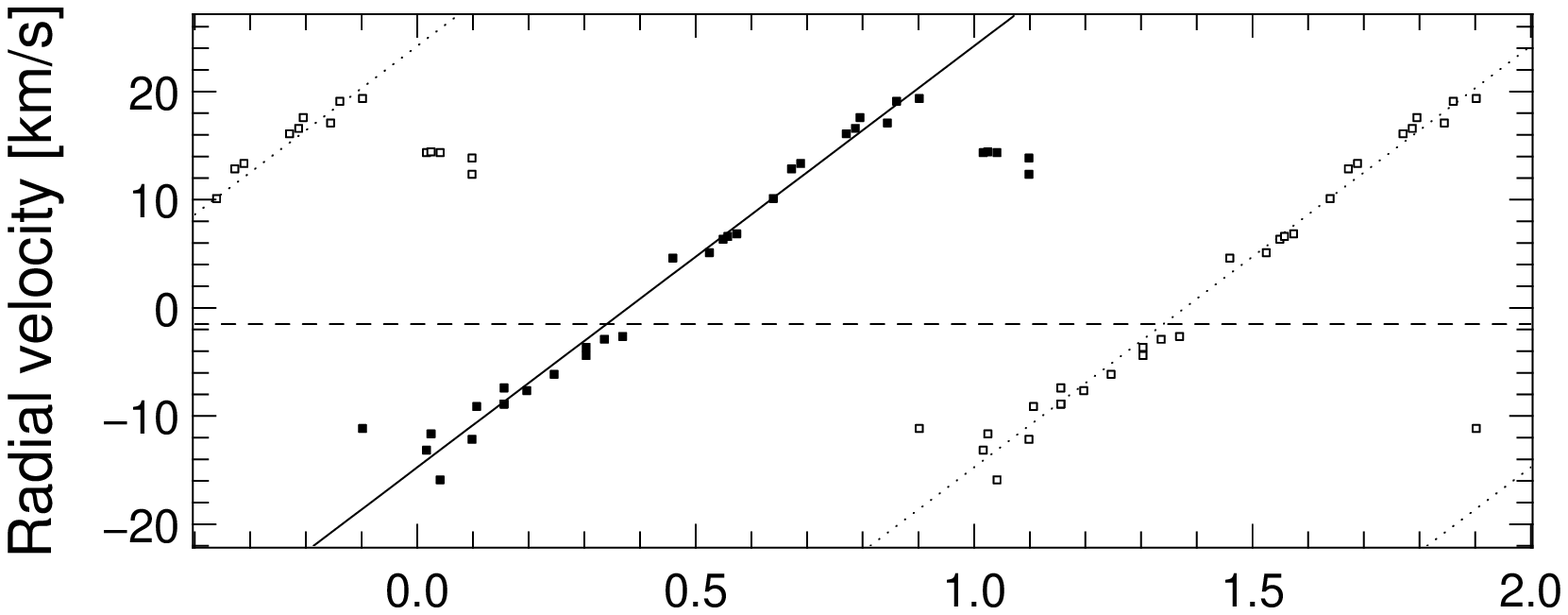} \\
\includegraphics[width=13cm]{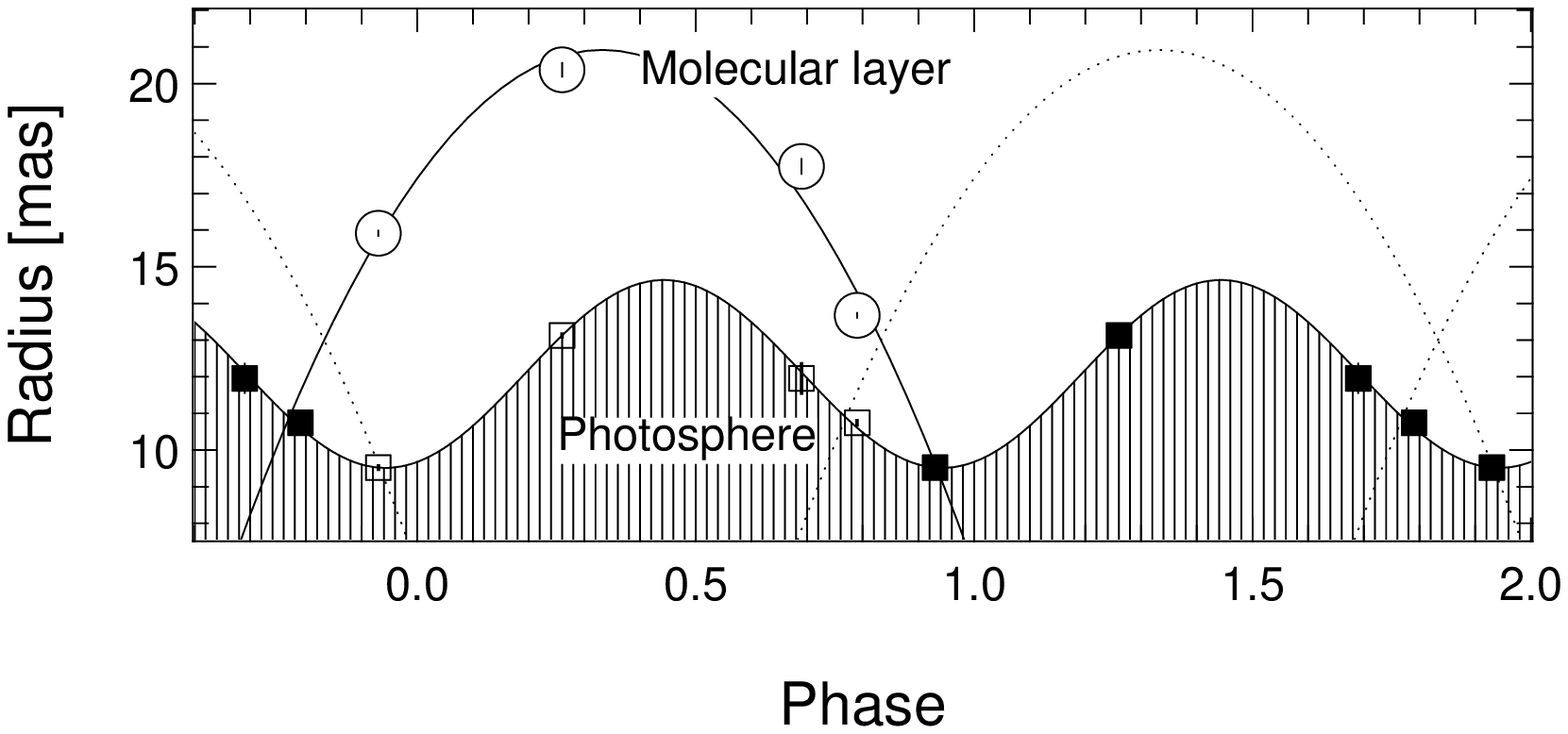}
\caption{{\it Upper panel}: Radial velocity of the CO ($\Delta v =3$)
  molecule from \citet{1982ApJ...252..697H}, folded in phase and
  shifted by 9.6 km/s \citep{1990ApJ...358..251W}. The tilted lines
  result from a linear fit over the $0<\phi<0.8$ period. It
  corresponds to a constant acceleration of
  $g_{\rm radial\ velocity}=-(1.10\pm0.04)$\,mm.s$^{-2}$. {\it Lower panel}: ballistic
  trajectory of inward acceleration fitted to the position of the
  molecular layer. The highest point of the ballistic trajectory is at
  $\phi=0.36$.}
\label{fig:dynam}
\end{figure}

In the upper panel of Fig.~\ref{fig:dynam} are plotted radial velocity
data from \citet{1982ApJ...252..697H}. For convenience, the data have
been folded in phase, and shifted by 9.6 km/s. This velocity shift
comes from an estimation of the local stellar velocity from CO
measurements by \citet{1990ApJ...358..251W}, but is not relevant to
derive the acceleration.  We performed a linear fit over the period
($0<\phi<0.8$) and derived a mean acceleration $g_{\rm
  radial\ velocity}=-(1.10\pm0.04$)\,mm.s$^{-2}$.  Simultaneously, we
fitted a parabola of constant inward acceleration over the position of
the molecular layer (lower panel of Fig.~\ref{fig:dynam}). According
to our measurements, the angular acceleration of the layer is $g_{\rm
  angular}=-(5.4\pm1.4)\times10^{-14}$\,mas.s$^{-2}$.  Using
Eq.~(\ref{eq:para}), this gives a parallax of $5.9\pm1.5$\,mas.  The
main source of uncertainty is coming from the estimation of the
geometric acceleration of the layer, $g_{\rm angular}$.

Several values of $\chi$ Cyg's parallax exist in the literature, but
no consensus exists yet, due to the difficulty of measuring the
position of such a star of high angular diameter and inhomogeneous
surface brightness. We can cite, among others, a measurement by the
Allegheny Observatory \citep[$8.8\pm1.9$\,mas,
  in][]{1991ApJ...377..669S}, and multiple parallax calculations
obtained from the Hipparcos dataset: $9.43\pm1.36$\,mas in
\citet{1997A&A...323L..49P}, $6.71\pm1.00$\,mas in
\citet{2003A&A...403..993K}, and $5.71\pm1.12$\,mas in
\citet{leeuwen07}. Our measurement is compatible with this most recent
value.  Another way to measure the distance is to use the
Period/Luminosity relation of Mira variables.
According to Eq.~(\ref{eq:PL}), the absolute magnitude of $\chi$ Cyg
should be $M_k=-8.06\pm0.09$.  Assuming a K magnitude of $-1.95$
\citep[including interstellar or circumstellar extinction
  from][]{2003A&A...403..993K}, the distance of $\chi$ Cyg should be
6.1\,mas. This last value is also compatible with our measurement.

\subsection{Mass derivation}
\subsubsection{The gravitational field}

Assuming a free-falling molecular layer, the mass of the star can be
derived by:
\begin{equation}
M_\star=\cfrac{g\,R_{\rm layer}^2}{G}\,,
\end{equation}
where $g=g_{\rm radial\ velocity}/p$, $G$ the universal gravitational
constant, and $R_{\rm layer}$ the position of the molecular
layer. Over the period $0<\phi<0.8$, the average radius of the
molecular layer is 18\,mas, which gives $R_{\rm
  layer}=3.0^{+1.0}_{-0.6}\,$AU (for a parallax of
$5.9\pm1.5\,$mas). Assuming a gravitational acceleration of
$g=-(1.37\pm0.05$)\,mm.s$^{-2}$, this leads to a stellar mass
$M_\star=2.1^{+1.5}_{-0.7}\,M_\sun$. Note that the uncertainty on the
parallax dominates the error bar.

We are fully aware that the free-fall approximation is an important
assumption. Three other forces can affect the trajectory of the
molecular layer:
\begin{itemize}
\item pressure force
\item radiative pressure
\item centripetal force due to the rotation of the star
\end{itemize}
The question is whether we can neglect them compared to the
gravitational force. If not, our estimate would only yield a lower
limit of the mass.

Centripetal force could be dismissed due to the long rotation period
of this evolved star. From conservation of momentum energy, it is
likely that the star has a period well over several tens of years.

Radiative pressure is responsible for the large mass loss. However,
because of the small cross-section between CO and starlight, the
kinetic energy transferred to the molecules would not result in a
sensible acceleration. In other words, we expect $\beta_{{\rm
    CO}\,(\Delta v =3)}$ the ratio of radiation pressure to stellar
gravity to be significantly below 1.  A dragging effect could be
envisioned, but the velocity gradient between grains and molecules is
necessarily small so low in the atmosphere.

Gas pressure force in-between two shocks is likely to be small
\citep{2000ARA&A..38..573W}. This can be understood as follows: the
density gradient of simulated static atmosphere of Mira stars are
extremely steep \citep{1988ApJ...329..299B}. In a static atmosphere,
pressure force induced by density gradient are rigorously equal to the
gravitational force. In pulsating Mira stars, the atmosphere is
``puffed up'', with a density gradient several orders of magnitude
smaller. Likewise, pressure forces will be several orders of magnitude
lower than the gravitational force (up to a certain distance from the
star).

From an observational point of view, this idea is mostly confirmed by
the constant acceleration of the CO ($\Delta v =3$)
molecules. Pressure forces and radiative pressure cannot be constant
over the pulsation cycle. Bolometric flux varies by at least a factor
2, and pressure forces may vary even more especially in the case of
multiple ballistic trajectories per cycle as proposed by
\citet{2006JAVSO..35...62W}.  Thus, if theses forces were dominant, we
would detect a non-linear effect on the acceleration of the
molecules. Since the radial velocity measurements of
\citet{1982ApJ...252..697H} rule out a difference in acceleration of
more than 10\% between phase 0 and phase 0.8, it is unlikely that
pressures forces slow the inward acceleration by more than a few
percents.

\subsubsection{The Period/Mass/Radius relation}

Indirect confirmation of $\chi$ Cyg mass is difficult. Indeed,
$2.1\,M_\sun$ is quite massive for a Mira star
\citep{1993ApJ...413..641V}. However, $\chi$ Cyg may not be a Mira of
the most common sort. First, the star has often been noticed by the
remarkable strength of his emission lines
\citep[eg][]{1947ApJ...106..274M}. Secondly, its 408 day period is
slightly longer then the average pulsation period of Mira stars
\citep{2000MNRAS.319..728W}. Lastly, according to our parallax
measurement, its mean radius of $440^{+150}_{-50}\,R_\sun$ (12.1\,mas)
makes $\chi$ Cygni a quite huge object.

The feeling that $\chi$ Cygni can be a quite massive Mira star is
confirmed by the P/M/R relation. Indeed, applying $\chi$ Cyg's
parameters to Eq.~(\ref{eq:PMR}) gives a mass of
$3.1^{+2.7}_{-1.2}\,M_\sun$. This value is compatible with our
measurement.

\subsection{Discussion} \label{sc:dicuss}

This work makes use of several assumptions which will certainly need
to be fine-tuned.  If anything, they emphasize the need to complete
the method with detailed numerical modeling to assess the influence of
atmospheric pressure forces. In the case of $\chi$ Cyg, the
consistency of the acceleration was the main argument to neglect
pressure effects. We also neglected the influence of an eventual
centrifugal force as it is unlikely because of the slow rotation of
the star's atmosphere. Finally, we assumed the molecular layer
detected in the interferometric data to be at the same position as the
CO ($\Delta v=3$) molecules observed by
\citet{1982ApJ...252..697H}. This last point may be the most arguable
one. Even though both spectroscopic and interferometric observations
have been made in the H band, it is true that contamination may exist
due to molecules (e.g. H$_2$O) possibly formed at a slightly different
height inside the atmosphere. Clear distinction between several
possible molecular layers will nevertheless be possible in the near
future thanks to high resolution spectro-interferometry. By probing
the apparent stellar diameter within the depth of saturated absorption
lines, it would provide the velocity and the position for individual
species inside the atmosphere. This is a prospect which would not only
strengthen our mass and distance determination method, but also would
be bound to revolutionize our understanding of shock driven
atmospheres.

Evolved stars could prove to be precious secondary distance
indicators. Compared to Cepheid, the fact that these stars are large
and extremely bright in the infrared (where interstellar absorption is
minimal), would make them a precious secondary indicator. To allow
that, the P/L relation of LPV stars should be properly assessed. This
work is a first step in that direction. Moreover, this method has an
intrinsic advantage compared to the traditional Baade-Wesselink method
since the projection factor can be derived from the position of the
molecular layer (even though we agree that the projection factor is
not yet the limiting term in the final accuracy of this method).

Deriving the mass of these stars also opens an important prospect to
study stellar structure, chemistry, and mass loss. It is one of the
main parameters (with radius) to constrain stellar instabilities and
oscillation modes. This novel way to derive the mass is applicable to
a wide variety of pulsating objects. The requirement is to be able to
probe a portion of the atmosphere where the gravitational field is
significantly stronger than the pressure forces (conventional and
radiative). It only happens in shock driven atmospheres, but
nevertheless should be observable in numerous types of pulsating
stars. For example, recent results on Cepheids showed a faint upper
atmospheric molecular layer likely to be gravitationally unstable
\citep{2006A&A...453..155M}. Note that this technique does not require
a regular pulsating variable; it can also be applied to any irregular
or semiregular variables, featuring shock driven molecular layers.

\acknowledgements 
We acknowledge with thanks the variable star
observations from the AAVSO International Database contributed by
observers worldwide and used in this research.  

\bibliographystyle{apj}
\bibliography{Cygbib,apj-jour}

\clearpage

\begin{deluxetable}{lcrc}
\tablecolumns{4} \tablewidth{0pc} \tablecaption{Observation Log
\label{tb:Config}}
\tablehead{
Date (UT) & $\phi$ & Configuration \tablenotemark{a}  & Length \tablenotemark{b}  [m]}
\startdata
2005 May 25& 0.92 &	A15-B15-C0&21\\
2005 May 27& 0.93 &	A15-B15-C5&	 21\\
2005 May 31& 0.93 &	A25-B15-C10&	 29\\
2005 Jun 1& 0.94  &	A35-B15-C10&		37\\
\hline
2005 Oct 5, 6, 7& 0.25 &	A5-B5-C0&	7\\
2005 Oct 8, 9& 0.26 &	A5-B15-C0&	14\\
2005 Oct 10, 11& 0.26 &	A15-B15-C0&21	\\
2005 Oct 12, 13& 0.27 &	A25-B15-C0&29	\\
\hline
2006 Mar 29, 31& 0.68 &	A15-B15-C0&21	\\
2006 Apr 2& 0.69 &	A5-B5-C0&7		\\
2006 Apr 7& 0.70 &	A30-B15-C0&	33	\\
\hline
2006 May 11& 0.78&	A15-B5-C10&	15	\\
2006 May 12& 0.78&	A15-B5-C0&	15	\\
2006 May 13& 0.79&	A15-B15-C0&	21	\\
2006 May 14& 0.79&	A30-B15-C0&	33	\\
2006 May 15& 0.79&	A35-B15-C21&	38	\\
2006 May 16& 0.79&	A35-B15-C25&	38	\\
\enddata
\tablenotetext{a}{\ Interferometer configuration refers to the location,
  in meters, of telescopes A, B, and C on the North/East, South/East
  and North/East arms, respectively} 
\tablenotetext{b}{\ Length of
  maximum projected baseline}
\end{deluxetable}

\clearpage

\begin{deluxetable}{ccc}
\tablecolumns{3} \tablewidth{0pc} \tablecaption{Calibrators
\label{tb:calib}}
\tablehead{
Calibrator & Spectral Type & UD diameter }
\startdata
HD\,176670 &    K2.5\,III   & $2.330 \pm  0.026$ \\
HD\,180450 &    M0\,III     & $2.770  \pm 0.032$ \\
HD\,186619 &    M0\,IIIab   & $2.190  \pm 0.025$ \\
HD\,188149 &    K4\,III     & $1.490  \pm 0.020$ \\
HD\,197989 &    K0\,III     & $4.440  \pm 0.048$ \\
\enddata
\end{deluxetable}

\clearpage

\begin{deluxetable}{lcccc}
\tablecolumns{5} \tablewidth{0pc} \tablecaption{Best-fit parameters
\label{tb:Res_fit}}
\tablehead{
  & May/June 2005  & October 2005 & March/April 2006 & May 2006  \\
 & ($\phi = 0.93$) & ($\phi = 0.26$) & ($\phi = 0.69$) & ($\phi =
0.79$) }
\startdata
$\theta_\star$ (mas)&  $19.04 \pm 0.09$ & $26.25 \pm 0.08$& $23.97 \pm  0.80$&$21.49 \pm  0.11$\\
LD [$\alpha$]        & $1.34 \pm 0.05$ & $1.08 \pm 0.03$& $2.54 \pm 0.39$&$2.42 \pm 0.05$ \\
$\theta_{\rm layer}$ (mas)& $31.83 \pm 0.15$ &  $40.75 \pm 0.37$& $35.48 \pm  0.40$&$27.35 \pm  0.13$\\
$F_{\rm layer}/F_{\rm total}$ (\%)& $6.5 \pm 0.2 $ & $4.7 \pm 0.2 $& $8.8 \pm  0.3$&$8.13 \pm  0.2$\\
$X_{\rm spot}$ (mas) & $5.22 \pm 0.05$& $-8.92 \pm 0.39$& $2.22 \pm  0.42$&$3.21 \pm  0.20$\\
$Y_{\rm spot}$ (mas) & $2.97 \pm 0.05$& $2.96 \pm 0.10$& $-4.24 \pm  0.34$&$-6.70 \pm  0.09$\\
$F_{\rm spot}/F_{\rm total}$ (\%) & $5.9 \pm 0.1 $&  $1.7 \pm 0.1 $& $3.7 \pm  0.3$&$1.7 \pm  0.1$\\
\hline
Reduced $\chi^2$   & 6.8 & 10.3 & 1.33 & 19.6\\
\enddata
\end{deluxetable}

\clearpage

\begin{deluxetable}{lcccccc}
\tabletypesize{\footnotesize}
\tablecolumns{7} \tablewidth{0pc} \tablecaption{Model dependent diameter measurements
\label{tb:fit}}
\tablehead{
  & \multicolumn{3}{c}{May 2005 ($\phi = 0.93$)} &
  \multicolumn{3}{c}{October 2005 ($\phi = 0.26$)}  \\
\hline
Model & UD  & UD + layer + spot & LD + layer + spot & UD & UD + layer
+ spot & LD + layer + spot }
\startdata
$\theta_\star$ (mas)& $16.24 \pm 0.07$ & $14.99 \pm 0.07$ & $19.04 \pm 0.09$ 
              & $22.99 \pm 0.11$ & $20.90 \pm 0.12$ & $26.25 \pm 0.08$
\\
$\alpha$      &  --              & --               & $1.35 \pm 0.05$ 
              & --               & --               & $1.08 \pm 0.03$
 \\
$\theta_{\rm layer}$ (mas)& --        & $21.76 \pm 0.27$ & $31.85 \pm 0.15$ 
              &  --              & $26.76 \pm 0.24$ & $40.75 \pm 0.37$
\\
$F_{\rm layer}/F_{\rm TOTAL}$ (\%)&-- & $10.9 \pm 0.3$& $6.5 \pm 0.2$ 
              & --               & $15.1 \pm 0.6$& $4.7 \pm 0.2$
\\
$X_{\rm spot}$ (mas)&        --       & $5.47 \pm 0.14$ & $5.22 \pm 0.05$
              & --               & $-8.4 \pm 0.78$   & $-8.92 \pm 0.39$
\\
$Y_{\rm spot}$ (mas)&  --             & $3.26 \pm 0.11$  & $2.97 \pm 0.05$
              & --               & $2.15 \pm 0.17$  & $2.96 \pm 0.10$
\\
$F_{\rm spot}/F_{\rm TOTAL}$ (\%)&-- & $5.9  \pm 0.2$& $5.9 \pm 0.1$
              & --               & $1.2  \pm 0.1$& $1.7 \pm 0.1$
\\
\hline
Reduced $\chi^2$ & 550           &   29          & 6  
         &        536            &   14          & 10
\\
\enddata
\end{deluxetable}

\clearpage

\begin{deluxetable}{lrrrr}
\tablecolumns{5} \tablewidth{0pc} \tablecaption{Bolometric Flux
\label{tb:flux}}
\tablehead{
$\phi$ & \multicolumn{1}{c}{0.96} & \multicolumn{1}{c}{0.26} & \multicolumn{1}{c}{0.69} & \multicolumn{1}{c}{0.79} \\
JD    & \multicolumn{1}{c}{2453518} & \multicolumn{1}{c}{2453653} & \multicolumn{1}{c}{2453826} & \multicolumn{1}{c}{2453869} }
\startdata
J (mag) \tablenotemark{a}& $0.00\pm0.15$  &   $-0.46\pm0.15$ & $0.15\pm0.15$ & $0.07\pm0.15$ \\
H (mag) \tablenotemark{a}& $-1.00\pm0.15$ &   $-1.65\pm0.15$ & $-1.05\pm0.15$ &  $-1.01\pm0.15$   \\
K (mag) \tablenotemark{a}& $-1.65\pm0.15$ &   $-2.24\pm0.15$ & $-1.73\pm0.15$  &  $-1.65\pm0.15$  \\
L (mag) \tablenotemark{a}& $-2.50\pm0.15$ &   $-2.84\pm0.15$ & $-2.51\pm0.15$  & $-2.48\pm0.15$  \\
$F_{\rm Bol}$ \\
(10$^{-13}$W\,cm$^{-2}$) & $6.83\pm0.38$ & $10.15\pm0.57$ & $6.80\pm0.35$ & $6.74\pm0.36$\\ 
\enddata
 \tablenotetext{a}{Magnitudes obtained from \citet{2000MNRAS.319..728W}}
\end{deluxetable}

\clearpage

\begin{deluxetable}{lcccc} 
\tablecolumns{5} \tablewidth{0pc} \tablecaption{Time dependent
  parameters of $\chi$ Cyg\label{tb:physics}} \tablehead{ & May/June
  2005 & October 2005 & March/April 2006 & May 2006 \\ 
\hline $\phi$ &
  \multicolumn{1}{c}{0.93} & \multicolumn{1}{c}{0.26} &
  \multicolumn{1}{c}{0.69} & \multicolumn{1}{c}{0.79} } 

\startdata
$\theta_\star$ (mas)&$19.04 \pm 0.09 $ & $ 26.25 \pm 0.08 $ & $ 23.97
\pm 0.80 $ & $21.49 \pm 0.11 $ \\ $R_\star/R_\sun$ \tablenotemark{a} &$348
\pm 94$ & $ 480 \pm 130$ & $ 439 \pm 119$ & $393 \pm 106$\\ $T_\star$
    (K)&$2742 \pm 45 $ & $ 2578 \pm 40 $ & $ 2441 \pm 72 $ & $2572 \pm
    41$ \\ \hline $\theta_{\rm layer}$ (mas)&$31.83 \pm 0.15 $ & $
    40.75 \pm 0.37 $ & $ 35.48 \pm 0.40 $ & $27.35 \pm 0.13 $
    \\ $R_{\rm layer}/R_\star$ &$1.67 \pm 0.02 $ & $ 1.55 \pm 0.02 $ &
    $ 1.48 \pm 0.07 $ & $1.27 \pm 0.01$ \\  $T_{\rm layer}$ (K) & $1821
    \pm 29 $ & $ 1795 \pm 28 $ & $ 1747 \pm 52 $ & $2032 \pm 32$
    \\ $\tau_{\rm layer}$ & $0.042 \pm 0.002 $ & $ 0.032 \pm 0.002 $ &
    $ 0.067 \pm 0.007 $ & $0.074 \pm 0.002$ \\ \hline $p$ \tablenotemark{b}
    & $1.22\pm0.01$&$1.26\pm0.01$&$1.22\pm0.02$&$1.27\pm0.01$
    \enddata
\tablenotetext{a}{Assuming a parallax of $5.9\pm1.5\,$mas (Sect.~\ref{sc:para})}
\tablenotetext{b}{Projection factor (Sect.~\ref{sc:p})}
\end{deluxetable}

\clearpage

\begin{deluxetable}{lrrr} 
\tablecolumns{4} \tablewidth{0pc} \tablecaption{Sinusoidal fit \tablenotemark{a}\label{tb:evolution}} \tablehead{ & \multicolumn{1}{c}{a} & \multicolumn{1}{c}{b} & \multicolumn{1}{c}{$\phi_0$} } 
\startdata
$F_{\rm Bol}$ ($10^{-13}\,$Wcm$^{-2}$) & $8.3\pm0.3$ & $1.8\pm0.4$ & $0.30 \pm0.03$\\
Diameter (mas) & $24.2\pm0.1$ & $5.1\pm0.1$ & $0.44\pm0.01$\\
Temperature (K)& $2560\pm30$ & $226\pm56$&$0.02\pm0.02$
    \enddata
\tablenotetext{a}{Sinusoidal defined by: $a+b\sin(2\pi(\phi-\phi_0))$}
\end{deluxetable}

\end{document}